\documentclass[reprint,
showpacs,
 amsmath,amssymb,
prb,
]{revtex4-1}

\usepackage{graphicx}
\usepackage{dcolumn}
\usepackage{longtable}
\usepackage{lipsum}
\usepackage{color}
\begin{document}


\title{ Green function theory of dirty two-band superconductivity}
\author{Yasuhiro Asano$^{1,2,3}$}
\author{Alexander A.\ Golubov$^{3,4}$}
\affiliation{$^{1}$Department of Applied Physics,
Hokkaido University, Sapporo 060-8628, Japan\\
$^{2}$Center of Topological Science and Technology,
Hokkaido University, Sapporo 060-8628, Japan\\
$^{3}$Moscow Institute of Physics and Technology, 141700 Dolgoprudny, Russia\\
$^{4}$Faculty of Science and Technology and MESA+ Institute for Nanotechnology, 
University of Twente,
7500 AE Enschede, The Netherlands
}%

\date{\today}

\begin{abstract}
We study the effects of random nonmagnetic impurities on the superconducting transition temperature $T_c$
in a two-band superconductor, where we assume an equal-time spin-singlet $s$-wave pair 
potential in each conduction band and the hybridization between the two bands as well as the band asymmetry. 
In the clean limit, the phase of hybridization determines the stability of two states: called $s_{++}$ and $s_{+-}$. 
The interband impurity scatterings decrease $T_c$ of the two states exactly in the same manner
when time-reversal symmetry is preserved in the Hamiltonian. 
We find that a superconductor with larger hybridization shows more moderate suppression of $T_c$.
This effect can be explained by the presence of odd-frequency Cooper pairs which are generated by 
the band hybridization in the clean limit and are broken by impurities.
\end{abstract}


\maketitle

\section{Introduction}

As shown in historical literature~\cite{agd,abrikosov:jetp1959,anderson:jpcs1959}, the superconducting transition temperature $T_c$ 
of a conventional $s$-wave superconductor 
is insensitive to the concentration of nonmagnetic impurities. On the other hand, the impurity scatterings 
reduce $T_c$ of an unconventional superconductor characterized by such symmetry as $p$-wave or $d$-wave. 
 The unconventional pair potential changes its sign on the Fermi surface depending on momenta of a quasiparticle. 
A quasiparticle can detect the sign of the pair potential 
while it travels a certain distance freely from any scatterings. 
The superconducting the coherence length $\xi_0$ is the characteristic distance 
of such ballistic motion. Therefore, the mean free path $\ell$ due to elastic impurities
must be much longer than $\xi_0$ to realize unconventional superconductivity.

The robustness of $s$-wave superconductivity under impurity scatterings seems to be weakened  
in multiband superconductors such as heavy fermionic compounds~\cite{stewart:rmp2011},
 MgB$_2$~\cite{mgb2:akimitsu2001,mgb2:louie2002}, iron pnictides~\cite{pnictide:hosono2008}, 
 and Cu-doped Bi$_2$Se$_3$~\cite{hor:prl2010,fu:prl2010}.
To make the argument simple, let us consider a two-band superconductor~\cite{allen:review1982} in which
the $\lambda$ th conduction band has an $s$-wave spin-singlet pair potential $\Delta_\lambda$ for $\lambda=1-2$.
In pnictides, for instance, experimental results suggest a fully gapped superconducting order parameter.~\cite{hashimoto:prl2009,evtushinsky:njp2009,nakayama:epl2009}
In addition to a conventional $s$-wave state $\Delta_1  \Delta_2>0$ ($s_{++}$ state), 
theories\cite{mazin:prl2008,kuroki:prl2008,chubukov:prb2008,raghu:prb2008} have 
indicate a sign-changing superconducting order parameter with $\Delta_1  \Delta_2< 0$ ($s_{+-}$ state).

It has been well established that the interband impurity scatterings reduce $T_c$ in a multiband 
superconductor~\cite{golubov:prb1997,efremov:prb2011,korshunov:prb2014,korshunov:uspekhi2016,onari:prl2009,asano17interband,hoyer:prb2015}.
According to the existing theories~\cite{efremov:prb2011,onari:prl2009}, an $s_{+-}$ state
is more fragile than an $s_{++}$ state under potential disorder. 
The conclusion has been understood in terms of an analogy to the effects of impurities 
in unconventional superconductors.
Namely, the diffusive impurity scatterings wash out the sign difference between the two pair potentials.
It has been demonstrated that strong potential disorder causes
the transition from an $s_{+-}$ state to an $s_{++}$ state~\cite{efremov:prb2011} near $T_c$. 
In addition, the ground state in the presence of impurities breaks time-reversal symmetry 
spontaneously~\cite{stanev:prb2014}.
At present,  mechanisms for the time-reversal-symmetry-breaking state are an open question. 
We address this issue in the present paper.

A unique aspect of two-band superconductors might be the effects of band hybridization.
Black-Schaffer and Balatsky~\cite{BSchaffer:prb2013,BSchaffer:prb2013r} have shown that
the band hybridization generates odd-frequency pairs~\cite{berezinskii:jetplett1974} in 
the uniform ground state. Odd-frequency pairs exhibit a paramagnetic response 
to an external magnetic field~\cite{tanaka:prb2005,asano11,higashitani:prl2013,mironov:prl2012,fominov:prb2015,suzuki:prb2014,suzuki:prb2015}, 
which has been confirmed recently by a $\mu$SR measurement~\cite{dibernardo:prx2015}.
Odd-frequency pairs are thermodynamically unstable because of their paramagnetic property. 
Therefore, the presence of odd-frequency pairs reduces $T_c$ in a uniform two-band 
superconductor in the clean limit~\cite{asano15}. It has been unclear how odd-frequency pairs
modify $T_c$ in the presence of impurities.

In this paper, we first derive the mean-field Hamiltonian of a time-reversal two-band superconductor 
in the presence of hybridization between the two bands $v\, e^{i \theta}$ as well as the band 
asymmetry $\gamma$. 
We assume an equal-time spin-singlet $s$-wave pair 
potential in each conduction band $\Delta_\lambda=|\Delta_\lambda|e^{i\varphi_\lambda}$ for $\lambda=1-2$.
We will show that these phases in the Hamiltonian must satisfy $\exp \{i(2\theta-\varphi_1+\varphi_2)\}=1$ 
in order to preserve time-reversal symmetry of the Hamiltonian. 
Namely $\theta=0$ ($\theta=\pi/2$) favors an $s_{++}$ ($s_{+-}$) state.
Next, we study the effects of impurity scatterings on the transition temperature 
 on the basis of the standard Green function theory of superconductivity.
The effects of impurity scatterings are considered through the self-energy which is estimated within 
the Born approximation. The transition temperature is calculated by solving the gap equation. 
In contrast to the results in Ref.~\onlinecite{efremov:prb2011}, 
the interband impurity scatterings reduce $T_c$ exactly in the same manner in the two states (i.e., $s_{++}$ and $s_{+-}$ ).
The time-reversal symmetry of the Hamiltonian explains reasons of the 
discrepancy between the two theories. We will show that an $s_{++}$ state and an $s_{+-}$ state 
are unitary equivalent to each other and that the gap equations always gives time-reversal 
ground state as long as time-reversal symmetry is preserved in the Hamiltonian.
We discuss also how odd-frequency Cooper pairs modify $T_c$ in a two band superconductor.

This paper is organized as follows. In Sec.~II, we describe a time-reversal superconducting state in a two-band 
superconductor in terms of a microscopic Hamiltonian. The solution of the Gor'kov equation and an important property
of the gap equation are discussed in the clean limit.
In Sec.~III, we analyze the symmetry of Cooper pairs in a two-band superconductor.
The effects of impurity scatterings on $T_c$ are studied by calculating the self-energy 
within the Born approximation in Sec.~IV. 
The relation between the results of the present paper and those in 
the previous paper~\cite{efremov:prb2011} is discussed in Sec.~V. 
The conclusion is given in Sec.~VI.
Throughout this paper, we use the units of $k_\mathrm{B}=c=\hbar=1$, where $k_{\mathrm{B}}$ 
is the Boltzmann constant and $c$ is the speed of light.

\section{Time-reversal two-band superconductor}
\subsection{Hamiltonian}\label{trs}

\begin{figure}[tbh]
\begin{center}
\includegraphics[width=8.5cm]{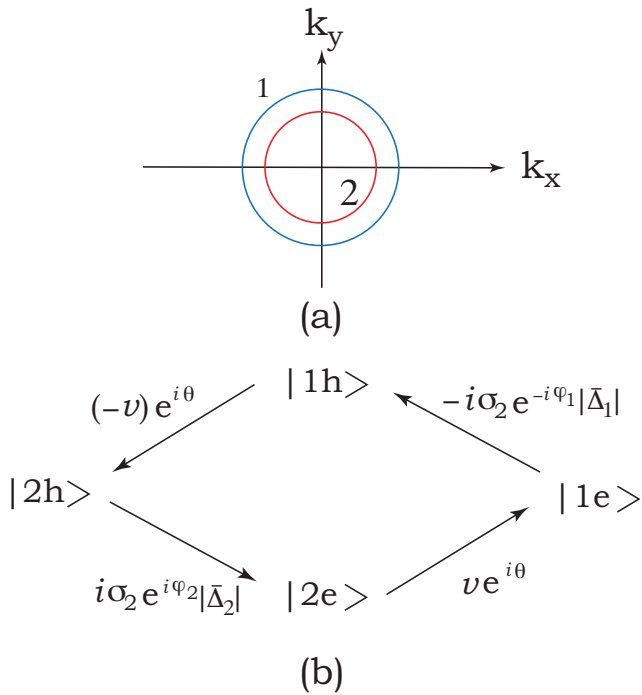}
\end{center}
\caption{
(a) The Fermi surfaces of the two bands are illustrated on two-dimensional momentum space, 
where 1 and 2 denote the circular Fermi surface of the first band and that of the second band, respectively.
(b) The matrix elements in Eq.~(\ref{hbdg_real}) which connect the particle state at the first band $|1e \rangle$ with
the hole state at the second band $|2h\rangle$. }
\label{fig:element}
\end{figure}

The Bogoliubov-de Gennes Hamiltonian can be described by $8 \times 8$ matrix form which reflects 
spin, particle-hole, and two-band degree of freedom. Let us define the Pauli matrices
 $\hat{\sigma}_j$ in spin space, $\hat{\rho}_j$ in two-band space and $\hat{\tau}_j$ in particle-hole 
 space for $j=1-3$. The unit matrix in these spaces are $\hat{\sigma}_0$, $\hat{\rho}_0$ and $\hat{\tau}_0$, 
 respectively.
The superconducting states of a two-band superconductor are described by
\begin{widetext}
\begin{align}
\mathcal{H}_{0} =& \int d\boldsymbol{r} 
\left[ \psi_{1,\uparrow}^\dagger(\boldsymbol{r}), \psi_{1,\downarrow}^\dagger(\boldsymbol{r}),
 \psi_{2,\uparrow}^\dagger(\boldsymbol{r}), \psi_{2,\downarrow}^\dagger(\boldsymbol{r}),
 \psi_{1,\uparrow}(\boldsymbol{r}), \psi_{1,\downarrow}(\boldsymbol{r}),
 \psi_{2,\uparrow}(\boldsymbol{r}), \psi_{2,\downarrow}(\boldsymbol{r})\right] \nonumber\\
&\times \check{H}_{0}
\left[
 \psi_{1,\uparrow}(\boldsymbol{r}), \psi_{1,\downarrow}(\boldsymbol{r}),
 \psi_{2,\uparrow}(\boldsymbol{r}), \psi_{2,\downarrow}(\boldsymbol{r}),
\psi_{1,\uparrow}^\dagger(\boldsymbol{r}), \psi_{1,\downarrow}^\dagger(\boldsymbol{r}),
 \psi_{2,\uparrow}^\dagger(\boldsymbol{r}), \psi_{2,\downarrow}^\dagger(\boldsymbol{r}) 
 \right]^{\mathrm{T}},\\
 \bar{H}_{0}(\theta, \varphi_1,\varphi_2)=& \left[ \begin{array}{cccc} 
 \xi_1(\boldsymbol{r})\hat{\sigma}_0 
  & ve^{i\theta} \hat{\sigma}_0 & |\bar{\Delta}_1| e^{i\varphi_1} i\hat{\sigma}_2 & 0\\
 ve^{-i\theta} \hat{\sigma}_0 & \xi_2(\boldsymbol{r}) \hat{\sigma}_0& 0 &|\bar{\Delta}_2| e^{i\varphi_2} i\hat{\sigma}_2\\
 -|\bar{\Delta}_1| e^{-i\varphi_1} i\hat{\sigma}_2 & 0 & -\xi_1(\boldsymbol{r}) \hat{\sigma}_0 & 
 -ve^{-i\theta} \hat{\sigma}_0\\
0&  -|\bar{\Delta}_2| e^{-i\varphi_2} i\hat{\sigma}_2 & -ve^{i\theta} \hat{\sigma}_0 & -\xi_2(\boldsymbol{r}) \hat{\sigma}_0
\end{array}\right], \label{hbdg_real}\\
\bar{\Delta}_1 =& \Delta_1 + \frac{g_{12}}{g_2} \Delta_2=|\bar{\Delta}_1|e^{i\varphi_1},\quad
\bar{\Delta}_2 = \Delta_2 + \frac{g_{12}^\ast}{g_1} \Delta_1=|\bar{\Delta}_2|e^{i\varphi_2}. \label{bd_def}
\end{align}
\end{widetext}
where $\psi_{\lambda,\sigma}^\dagger(\boldsymbol{r})$ ($\psi_{\lambda,\sigma}(\boldsymbol{r})$) is the  
creation (annihilation) operator of an electron with spin $\sigma$ ($=\uparrow$ or $\downarrow$) at the $\lambda$ th 
conduction band,
 $\xi_\lambda(\boldsymbol{r}) = - \nabla^2/(2m_\lambda) + \epsilon_\lambda - \mu_F$ is the kinetic 
energy at the $\lambda$ th band, $ve^{i\theta}$ denotes the hybridization between the two bands, 
and $\mathrm{T}$ means the transpose of a matrix. In Fig.~\ref{fig:element}(a), we schematically illustrate 
the Fermi surfaces of the two bands on two-dimensional momentum space. 
We assume a uniform spin-singlet $s$-wave pair potential for each conduction band which is defined by
\begin{align}
\Delta_\lambda = g_\lambda \left\langle \psi_{\lambda, \uparrow}(\boldsymbol{r})  
\psi_{\lambda, \downarrow}(\boldsymbol{r}) \right\rangle, \label{delta_def}
\end{align}
where $g_\lambda>0$ represents the attractive interaction between two electrons at the $\lambda$ th band.
Within the mean-field theory, the attractive interaction couples also 
$\Delta_1$ and $\Delta_2$ as shown in Eq.~(\ref{bd_def}), where $g_{12}$ represents such 
interband interaction
and its amplitude is considered to be smaller than $g_\lambda$, (i.e., $|g_{12}| < g_\lambda$). 
The details of the derivation are given in Appendix~\ref{model}. 
To discuss time-reversal symmetry of the Hamiltonian, we represents the phase of pair 
potential $e^{i\varphi_\lambda}$ explicitly.
Generally speaking, $g_{12}$ can be a complex number as well as the hybridization.
These phases are originated from the relative phase of the atomic orbital 
functions as shown in Appendix~\ref{model}. 
When we set the phase of the hybridization as $v e^{i\theta}$, we must choose 
the phase of the interband interaction as $g_{12}=|g_{12}| e^{2i\theta}$ to keep the consistency of the theory.

Time-reversal symmetry of a Hamiltonian $\bar{H}_0$ is represented by
\begin{align}
&\bar{\mathcal{T}}\, \bar{H}_0\, \bar{\mathcal{T}}^{-1} =\bar{H}_0,\\
&\bar{\mathcal{T}} = i\hat{\sigma}_2\, \hat{\rho}_0\, \hat{\tau}_0\, \mathcal{K}, \quad \mathcal{T}^2=-1,
\end{align}
where $\mathcal{K}$ means the complex conjugation. 
The single-particle Hamiltonian in Eq.~(\ref{hbdg_real}) does not contain either spin-dependent potentials or
vector potentials.
Thus, it is possible to show time-reversal symmetry of $\bar{H}_0$ if we find 
a unitary transformation that eliminates all the phase factors in Eq.~(\ref{hbdg_real}).
By applying 
the unitary transformation,  
\begin{align}
\bar{U}_{\varphi} &= \textrm{diag}[e^{i\frac{\varphi_1}{2}}\hat{\sigma}_0, e^{i\frac{\varphi_2}{2}}\hat{\sigma}_0, 
e^{-i\frac{\varphi_1}{2}}\hat{\sigma}_0, e^{-i\frac{\varphi_2}{2}}\hat{\sigma}_0] \\
&= \frac{\hat{\rho}_0+\hat{\rho}_3}{2}e^{i\frac{\varphi_1}{2}\hat{\tau}_3}
+\frac{\hat{\rho}_0-\hat{\rho}_3}{2}e^{i\frac{\varphi_2}{2}\hat{\tau}_3}, 
\end{align}
the Hamiltonian is transformed into
\begin{align}
&\bar{U}_{\varphi}^\dagger\, \bar{H}_{0}(\theta,\varphi_1,\varphi_2)\, \bar{U}_{\varphi} =
\bar{H}_{0}\left(\theta-\frac{\varphi_1-\varphi_2}{2},0,0\right).
\end{align}
Therefore, Eq.~(\ref{hbdg_real}) preserves time-reversal symmetry when
\begin{align}
2\theta - \varphi_1 + \varphi_2=2\pi n, \label{trs_cond}
\end{align}
is satisfied.
The phases of the two pair potentials and that of the hybridization are linked to 
one another when the superconductor preserves time-reversal symmetry.

The condition in Eq.~(\ref{trs_cond}) can be interpreted as follows. There are two routes which connect 
the particle states at the first band $|1e\rangle$ with the hole state at the second band $|2h\rangle$ 
as shown in Fig.~\ref{fig:element}(b). In the top route, $|1e\rangle$ first transits to $|1h\rangle$ by 
the pair potential $-|\bar{\Delta}_1|e^{-i\varphi_1} i \hat{\sigma}_2$ then reaches $|2h\rangle$ 
by the hybridization $-ve^{i\theta}$. 
The return process goes through $|2e\rangle$ as shown in the bottom route.
Namely $|2h\rangle$ first transits to $|2e \rangle$ by the pair potential $|\bar{\Delta}_2|e^{i\varphi_2} i\hat{\sigma}_2$
then returns back to $|1e\rangle$ by the hybridization $v e^{i\theta}$.
The matrix elements in the scattering processes becomes $ -v^2|\bar{\Delta}_1| |\bar{\Delta}_2| e^{i(2\theta-\varphi_1+\varphi_2)}$.
The factor $-1$ is derived from the particle-hole transformation between 
the single-particle hamiltonian in the electron branch $H_{e}$ and that of the hole branch $H_{h}$
because they are related to each other by $H_h=-H_e^\ast$.
The remaining factor $e^{i(2\theta-\varphi_1+\varphi_2)}$ plays a role of \textsl{magnetic flux} in two-band space.
Thus Eq.~(\ref{trs_cond}) must be necessary so that the Hamiltonian $\bar{H}_0$ preserves time-reversal symmetry.
As we will discuss in the next subsection,
the solutions of the gap equation always meet the condition in Eq.~(\ref{trs_cond}).

\subsection{Solution of Gor'kov equation}\label{gorkov}
 The Hamiltonian in Eq.~(\ref{hbdg_real}) in the momentum representation $\bar{H}_{0}(\boldsymbol{k})$
has the energy eigenvalues
\begin{align}
E_\pm^2 =& \xi_+^2+\xi_-^2 + |\Delta_+|^2 + |\Delta_-|^2 + v^2 \pm  
2 \sqrt{ Y}, \label{energy_app}
\end{align}
where we have defined 
\begin{align}
\xi_\pm=&\frac{\xi_{1,\boldsymbol{k}} \pm \xi_{2,\boldsymbol{k}} }{2}, \quad
\Delta_\pm=\frac{\bar{\Delta}_{1} \pm \bar{\Delta}_{2} }{2},\\
ve^{i\theta}=&v_1+iv_2 \quad K= \xi_+ \xi_- +{D_+},
\end{align}
\begin{align}
D_\pm=& \frac{\Delta_+ \Delta_-^\ast \pm \Delta_+^\ast \Delta_-}{2},\\
Y=& K^2 + v^2\xi_+^2 +v_1^2|\Delta_-|^2 + v_2^2|\Delta_+|^2\nonumber\\
&- i2 v_1v_2 D_-.
\end{align}

The Green function is obtained by solving the Gor'kov equation,
\begin{align}
&\left[ i\omega_n \bar{1}- \bar{H}_{0}(\boldsymbol{k}) \right] \bar{\mathcal{G}}_0(\boldsymbol{k},i\omega_n)=\bar{1},\\
& \bar{\mathcal{G}}_0(\boldsymbol{k},i\omega_n)
=\left[ \begin{array}{cc}
\check{\mathcal{G}}_0(\boldsymbol{k},i\omega_n) & \check{\mathcal{F}}_0(\boldsymbol{k},i\omega_n) \\
-\check{\mathcal{F}}_0^\ast(-\boldsymbol{k},i\omega_n) & -\check{\mathcal{G}}_0^\ast(-\boldsymbol{k},i\omega_n) 
\end{array}\right],
\end{align}
where $\omega_n=(2n+1)\pi T$ is a fermionic Matsubara frequency with $T$ being a temperature.
We find the exact solutions as~\cite{asano15}
\begin{widetext}
\begin{align}
\check{\mathcal{G}}_0(\boldsymbol{k},i\omega_n)
=&\frac{\hat{\sigma}_0}{2Z} 
\left[
\left\{ (- i\omega_n - \xi_+ ) X + v^2 \xi_+  +  \xi_- K \right\} \hat{\rho}_0 
+ \left\{ (- X +  \xi_+^2 + |\Delta_-|^2 +  i \omega_n \xi_+ )v_1 -iD_- v_2 \right\} \hat{\rho}_1 \right. \nonumber \\
&+ \left\{ (- X -  \xi_+^2 - |\Delta_+|^2 -  i \omega_n \xi_+)iv_2 + D_- v_1 \right\} i\hat{\rho}_2 
\left.
+ \left\{ -X \xi_- + ( i\omega_n + \xi_+ ) K \right\} \hat{\rho}_3 \right],\label{g_tra}\\
%
%
\check{\mathcal{F}}_0(\boldsymbol{k},i\omega_n)
=&\frac{i\hat{\sigma}_2}{2Z} 
\left[
\left\{ (-X +v_2^2) \Delta_+ +  ( K  + i v_1v_2) \Delta_-  \right\} \hat{\rho}_0
\right. 
+ \left\{ v_1( \xi_+ \Delta_+ -\xi_-\Delta_- ) -iv_2 ( \xi_+ \Delta_- -\xi_-\Delta_+ ) \right\}
\hat{\rho}_1 \nonumber\\
&+\omega_n ( v_1  \Delta_- -i v_2 \Delta_+ )  \hat{\rho}_2 
+  \left.  \left\{ (-X +v_1^2)  \Delta_- + ( K -iv_1v_2) \Delta_+  \right\} 
\hat{\rho}_3   \right],\label{f_tra}
%
\end{align}
\end{widetext}
\begin{align}
Z=& X^2 - Y,\\
X =& \frac{1}{2}[\omega_n^2 + \xi_+^2 + \xi_-^2 + |\Delta_+|^2 + |\Delta_-|^2 + v^2],
\end{align}
The diagonal elements of the anomalous Green function in band space are linked to the pair potentials in Eq.~(\ref{delta_def}), 
\begin{align}
\Delta_\lambda=&- g_\lambda T\sum_{\omega_n} \frac{1}{V_\mathrm{vol}} \sum_{\boldsymbol{k}} \nonumber\\
\times& \mathrm{Tr} \left[ \frac{\hat{\rho}_0 + s_\lambda \hat{\rho}_3}{2} \, 
 \check{\mathcal{F}}_0(\boldsymbol{k},i\omega_n) \, \frac{(-i\hat{\sigma}_2)}{2}\right], \label{scf_g}
\end{align}
where $s_\lambda$ is 1 \, ($-1$) for $\lambda=1 \, (2)$.
Together with $g_{12}=|g_{12}|e^{2i\theta}$, the self-consistent equation for the pair potential in Eq.~(\ref{bd_def})
 becomes
\begin{align}
&\bar{\Delta}_\lambda= T\sum_{\omega_n}\frac{1}{V_{\mathrm{vol}}}\sum_{\boldsymbol{k}}\frac{1}{4Z}\nonumber\\
&\times
\left[ \left\{ g_\lambda( \omega_n^2 + \xi_{\bar{\lambda}}^2+|\bar{\Delta}_{\bar{\lambda}}|^2) 
+|g_{12}|v^2 \right\} \bar{\Delta}_\lambda 
\right. \nonumber\\
&\left.
+\left\{ |g_{12}|(\omega_n^2+\xi_{\lambda}^2+|\bar{\Delta}_\lambda|^2) 
+g_\lambda v^2 \right\} e^{2i\theta\, s_{\lambda} } \bar{\Delta}_{\bar{\lambda}} \right], \label{scfg}
%
\end{align}
where we define $\bar{\lambda}=2 \, (1)$ for $ \lambda=1 \, (2)$.
By representing the phase of the pair potential explicitly as 
$\bar{\Delta}_\lambda=|\bar{\Delta}_\lambda|e^{i\varphi_\lambda}$,
we find an important fact that the gap equation always gives the solution which satisfies 
the relation $2\theta-\varphi_1+\varphi_2=2\pi n$ automatically.
We have to pay attention to this point when we introduce the impurity potential 
which hybridizes the two bands in Sec.~\ref{impurity}.

When the relation Eq.~(\ref{trs_cond}) is satisfied,
the energy eigen value in Eq.~(\ref{energy_app}) and $Z$ are independent of $\theta$, $\varphi_1$ and $\varphi_2$.
In such case, it is possible to define the unitary transformation which connects all the 
Hamiltonians satisfying $2\theta-\varphi_1+\varphi_2=2\pi n$.

\section{Odd-frequency Cooper pair}
In what follows, we represents the Hamiltonian in a reduced $4\times 4$ structure by choosing 
spin of an electron $\uparrow$ and spin of a hole $\downarrow$.
We assume that $\xi_\lambda(\boldsymbol{k}) = \xi(\boldsymbol{k}) - s_\lambda \gamma$ with 
$\xi(\boldsymbol{k})= \boldsymbol{k}^2/(2m) - \mu_F$, where $\gamma$ 
represents the band asymmetry. We also assume that $\gamma$ is much smaller than $\mu_F$.
The Hamiltonian is represented in $4 \times 4$ matrix form 
%
\begin{align}
\check{H}_0 =&\left[ \begin{array}{cccc}
\xi_{\boldsymbol{r}} -\gamma & ve^{i\theta}  & \bar{\Delta}_1 & 0 \\
ve^{-i\theta} & \xi_{\boldsymbol{r}} +\gamma & 0 & \bar{\Delta}_2 \\
\bar{\Delta}_1^\ast & 0 & -\xi_{\boldsymbol{r}}+\gamma & -ve^{-i\theta}  \\
0 & \bar{\Delta}_2^\ast & -ve^{i\theta}  & -\xi_{\boldsymbol{r}}- \gamma
\end{array}\right]. \label{h0}
%
\end{align}
In this section and Sec.~\ref{impurity}, we set $\theta=0$ for simplicity.
In what follows, we discuss the gap equation 
within the first order of $\bar{\Delta}_\lambda$ because
$\bar{\Delta}_\lambda$ is much smaller than another energy scales
near the transition temperature $T \lesssim T_c$.
The normal Green function in the linear regime becomes
\begin{align}
&\hat{\mathcal{G}}_0(\boldsymbol{k},\omega_n)= \frac{1}{Z_0}\nonumber\\
&\times\left[ 
\left\{ -(i\omega_n+\xi)(A_0+\xi^2) + 2(v^2 +  \gamma^2)\, \xi \right\}\hat{\rho}_0 
\right.\nonumber\\
&
-(A_0 - \xi^2 -2i\omega_n \xi)v \hat{\rho}_1
 \nonumber\\
& \left.+\left\{ A_0 + \xi^2 -2  \xi (i\omega_n+\xi) \right\} \gamma \hat{\rho}_3 \right].
\end{align}
The the anomalous Green function in the linear regime is also given by 
$\hat{\mathcal{F}}_0(\boldsymbol{k},\omega_n)= \sum_{\nu=0}^{3}{f}_\nu \hat{\rho}_\nu$ with
\begin{align}
f_0=&\frac{1}{Z_0} \left[-(A_0+\xi^2 ) \Delta_+   -2 \xi\, \gamma \Delta_-\right],\label{f0}\\
f_1= & \frac{1}{Z_0} \left[2 \, v \xi\, \Delta_+ + 2\, v \, \gamma \, \Delta_-\right],\label{f1}\\  
 f_2= & \frac{1}{Z_0} \left[ 2\,  \omega_n\,  v \, \Delta_-\right],\label{f2}\\
f_3=& \frac{1}{Z_0}\left[ -  2\, \xi\,  \gamma\,  \Delta_+ - (A_0 + \xi^2 - 2v^2) \Delta_-\right],\label{f3}\\
A_0=&\omega_n^2 + \gamma^2 + v^2,\; \Delta_\pm= \frac{1}{2}( \bar{\Delta}_1 \pm \bar{\Delta}_2),\\
Z_0=& \xi^4 +2 \xi^2(\omega_n^2 - \gamma^2 - v^2) + A_0^2.
\end{align}
The gap equation for $\theta=0$ in the linear regime is represented by,
\begin{align}
\bar{\Delta}_\lambda = & T\sum_{\omega_n}^{\omega_c} \frac{\pi N_0}{|\omega_n|A_0}
\left[ 
\left\{ g_\lambda (\omega_n^2+\gamma^2) + \frac{v^2}{2}(g_\lambda +|g_{12}|) \right\}
 \bar{\Delta}_\lambda \right.\nonumber\\
&\left. +
\left\{ |g_{12}|(\omega_n^2+\gamma^2) + \frac{v^2}{2}(g_\lambda+|g_{12}|)  \right\}
 \bar{\Delta}_{\bar{\lambda}} 
\right],\label{scf0}
\end{align}
where $\omega_c$ is the cut-off energy and $N_0$ is
the density of states at the Fermi level per spin. 
The summation over $\boldsymbol{k}$ is carried out by using the relation in Eq.~(\ref{k-int}).
Since we fix $\theta=0$, the solution satisfies 
$\varphi_1=\varphi_2$ meaning $\bar{\Delta}_1\bar{\Delta}_2>0$ as already mentioned 
in Eq.~(\ref{scfg}). 

Before turning into the effects of impurity scatterings, the symmetry of the pairing 
correlations should be summarized. 
The diagonal components, $f_0$ and $f_3$, belong 
to even-frequency spin-singlet even-momentum-parity
even-band-parity (ESEE) symmetry class and are linked to the pair potential~\cite{BSchaffer:prb2013,asano15,vasenko:jpcm2017,vasenko:jetplett2017}. 
An off-diagonal correlation $f_1$ belongs also to the ESEE class.  
The remaining component $f_2$, however, 
belongs to odd-frequency spin-singlet even-momentum-parity odd-band-parity (OSEO) class~\cite{BSchaffer:prb2013,asano15}. 
The thermodynamical stability of a pairing correlation depends directly on its frequency symmetry~\cite{asano11,suzuki:prb2014,suzuki:prb2015}. 
The superconducting state is realized when $E=F_S-F_N<0$, where $F_N$ ($F_S$) is the free energy in the normal (superconducting) 
state. To decrease the free energy, the superconducting condensate 
keeps its phase coherence. Therefore, the diamagnetism is the most fundamental property of all superconductors.
In the mean-field theory of superconductivity, the magnetic response of superconducting states is
described by the Meissner kernel $Q$ which is the linear response coefficient connecting the electric current 
$\boldsymbol{j}$ and the 
vector potential $\boldsymbol{A}$ as $\boldsymbol{j}=- Q (e^2/m) \boldsymbol{A}$~\cite{agd}. 
Phenomenologically, $Q$ is often refereed to as \textsl{pair density}.  
The contribution of the anomalous Green function to the Meissner kernel is given by
\begin{align}
&Q_F= T\sum_{\omega_n} \frac{1}{V_{\mathrm{vol}}} \sum_{\boldsymbol{k}} \frac{1}{2}\mathrm{Tr}\;
 \hat{\mathcal{F}}_0(\boldsymbol{k},i\omega_n)\; 
\hat{\mathcal{F}}^\ast_0(-\boldsymbol{k},i\omega_n),\\
&\frac{1}{2}\mathrm{Tr} \hat{\mathcal{F}}_0
\hat{\mathcal{F}}_0^\ast= f_0 f_0^\ast + f_1 f_1^\ast - f_2 f_2^\ast + f_3 f_3^\ast.
\label{qf}
\end{align}
The third term in Eq.~(\ref{qf}) is negative because $\hat{\rho}_2$ is pure imaginary.  
The results show that even- (odd-) frequency Cooper pairs have positive (negative) pair density 
and enhance (suppress) the Meissner effect~\cite{asano15}. 
The presence of usual even-frequency Cooper pairs decreases the free-energy.
On the other hand, the presence of odd-frequency pairs 
increases the free-energy~\cite{suzuki:prb2014,suzuki:prb2015} because they are thermodynamically unstable.
In Eq.~(\ref{scf0}), for instance, it is possible 
show that the hybridization $v$ reduces $T_c$. 
The hybridization induces the two pairing correlations: 
even-frequency interband pairing correlation $f_1$ and odd-frequency interband correlation $f_2$. 
The appearance of odd-frequency correlation suppresses $T_c$ 
because of their paramagnetic property~\cite{asano15}.
In addition to this, at $\bar{\Delta}_1=\bar{\Delta}_2$, Eq.~(\ref{scf0}) also show that 
$T_c$ in the presence of $v$  remains unchanged from that at $v=0$. 
The odd-frequency pairing correlation $f_2$ is absent in this case. 
These are key properties for understanding the variation of the transition temperature $T_c$ in the presence of 
impurities.

\section{Effects of impurities}\label{impurity}
Let us add the impurity potential 
\begin{align}
\check{H}_{\mathrm{imp}} =&V_{\mathrm{imp}}(\boldsymbol{r})\left[ \begin{array}{cccc}
1 & e^{i\theta} & 0 & 0 \\
e^{-i\theta} & 1& 0 & 0 \\
0 & 0 & -1 & - e^{-i\theta} \\
0 & 0 & - e^{i\theta} & -1
\end{array}\right]. \label{vimp}
\end{align}
to $\check{H}_0$ in Eq.~(\ref{h0}). The total Hamiltonian is $\check{H}=\check{H}_0+\check{H}_{\mathrm{imp}}$.
We emphasis that the interband impurity potential must have the same phase factor as the hybridization.
Otherwise, time-reversal symmetry is broken in the combined Hamiltonian $\check{H}$.
%
%
%

We assume that the impurity potential satisfies the following properties,
\begin{align}
\overline{V_{\mathrm{imp}}(\boldsymbol{r})}=&0,\label{vimp_1}\\
\overline{ V_{\mathrm{imp}}(\boldsymbol{r}) V_{\mathrm{imp}}(\boldsymbol{r}') } =& n_{\mathrm{imp}} v_{\mathrm{imp}}^2
\delta(\boldsymbol{r}-\boldsymbol{r}'), \label{vimp_2}
\end{align}
where $\overline{\cdots}$ means the ensemble average, $v_{\mathrm{imp}}$ represents the strength of impurity 
potential, and $n_{\mathrm{imp}}$ is the impurity density. 
We also assume that the attractive electron-electron interactions are insensitive to 
the impurity potentials~\cite{anderson:jpcs1959}.
Since $\theta=0$, the interband impurity potential is proportional to $\hat{\rho}_1 \hat{\tau}_3$ in Eq.~(\ref{vimp}).

In the presence of the impurity potential, the Green function within the Born approximation obeys
\begin{align}
&\left[ i\omega_n - \check{H}_0(\boldsymbol{k}) - \check{\Sigma}_{\mathrm{imp}} 
 \right]\check{\mathcal{G}}(\boldsymbol{k},\omega_n) = \check{1}, \label{gorkov1}\\
&\check{\Sigma}_{\mathrm{imp}}= \check{\Sigma}_a + \check{\Sigma}_b.
\end{align}
The self-energy due to the impurity scatterings are represented within the Born approximation as,
\begin{align} 
\check{\Sigma}_a=&n_{\mathrm{imp}} v_{\textrm{imp}}^2 \,
\check{\tau}_3\,   \frac{1}{V_{\mathrm{vol}}} \sum_{\boldsymbol{k}} \check{\mathcal{G}}_0(\boldsymbol{k},\omega_n)
\, \check{\tau}_3,\\
 =& \frac{\pi N_0 n_{\mathrm{imp}} v_{\textrm{imp}}^2}{|\omega_n|}
\left[ \begin{array}{cc} 
- i\omega_n \hat{\rho}_0 & \hat{s}_{-} \\
\hat{s}^\ast_{-}
 & - i\omega_n \hat{\rho}_0 \end{array}\right],\label{sigma_a1} \\
\check{\Sigma}_b=&n_{\mathrm{imp}} v_{\mathrm{imp}}^2 \,
\check{\tau}_3\, \hat{\rho}_1\,  \frac{1}{V_{\mathrm{vol}}} \sum_{\boldsymbol{k}} \check{\mathcal{G}}_0(\boldsymbol{k},\omega_n)
\, \hat{\rho}_1\, \check{\tau}_3.\\
 =& \frac{\pi N_0 n_{\mathrm{imp}} v_{\textrm{imp}}^2}{|\omega_n|}
\left[ \begin{array}{cc} 
- i\omega_n \hat{\rho}_0 & \hat{s}_{+} \\
\hat{s}^\ast_{+}
 & - i\omega_n \hat{\rho}_0 \end{array}\right],
\label{sigma_b1}
\end{align}
\begin{align}
\hat{s}_{\pm}=& \Delta_+ \hat{\rho}_0 -
\frac{\Delta_-}{A_0}v \gamma \hat{\rho}_1  \pm \hat{s}_{a},\\
\hat{s}_{a}=& \frac{\Delta_-}{A_0}
\left[  \omega_n v \hat{\rho}_2
-(\omega_n^2+\gamma^2) \hat{\rho}_3\right], \label{sa_def}
\end{align}
where $\Sigma_a$ and $\Sigma_b$ are the self-energy due to the intraband and that of the interband impurity 
scatterings, respectively. (See Appendix~\ref{selfenergy} for the derivation.)
As we will show later, $\Sigma_a$ does not change $T_c$ of a two-band superconductor. 
Therefore it is convenient to describe the self-energy as
\begin{align}
\check{\Sigma}_{\mathrm{imp}} =& \frac{1}{2 \tau_{\mathrm{imp}}|\omega_n|}
\left[ \begin{array}{cc} 
- i\omega_n  & \hat{s}_{-} \\
\hat{s}^\ast_{-}
 & - i\omega_n  \end{array}\right] 
 \nonumber\\&
 +\frac{1}{2 \tau_{\mathrm{imp}}|\omega_n|}
\left[ \begin{array}{cc} 
0 & \hat{s}_{a}  \\
\hat{s}_{a}^\ast & 0 \end{array}\right],\label{self1}\\
\frac{1}{\tau_{\mathrm{imp}}} =& 2 \times 2 \pi N_0 n_{\mathrm{imp}} v_{\textrm{imp}}^2.
\end{align}
Some parts of $\Sigma_b$ can be embedded into the first term in Eq.~(\ref{self1}) which does not change $T_c$.
The remaining part as shown in the second term Eq.~(\ref{self1}) modifies $T_c$. 
The interband scatterings wash out asymmetry in the pair potentials at the two bands, which 
suppresses the pairing correlations proportional to $\Delta_-$ in $f_2$ and $f_3$.
By solving the Gor'kov equation in the presence of impurities, we obtain the anomalous 
Green function within the lowest order of $\Delta_\pm$ as 
$\hat{\mathcal{F}}(\boldsymbol{k},\omega_n)= \sum_{\nu=0}^{3} \tilde{f}_\nu \hat{\rho}_\nu$.
The results after carrying out the summation over $\boldsymbol{k}$ are expressed as,  
\begin{align}
\langle \tilde{f}_0\rangle =& \frac{\pi N_0}{|{\omega}_n|}( -\Delta_+),\label{ft0}\\
\langle \tilde{f}_3\rangle =& \frac{\pi N_0}{|{\omega}_n| {A}_0}
\left[
({\omega}_n^2+\gamma^2) \Delta_- + I \Delta_- \right],\label{ft3}\\
I=&\frac{1}{ 2 \tau_{\mathrm{imp}} \,  |\tilde{\omega}_n|\, \tilde{A} }
\left[ - v^2 {\omega}_n \tilde{\omega}_n  + ( \tilde{\omega}_n^2+\gamma^2)({\omega}_n^2+\gamma^2) \right], \label{i_def}\\
\langle \tilde{f}_\nu\rangle \equiv& \frac{1}{V_{\mathrm{vol}}}\sum_{\boldsymbol{k}} \tilde{f}_\nu, \quad
\tilde{A}= \tilde{\omega}_n^2 + v^2 +\gamma^2,\\
\tilde{\omega}_n=&\omega_n \eta_n, 
 \quad \eta_n=\left(1 +\frac{1}{ 2 \tau_{\mathrm{imp}}|\omega_n| } \right),
\end{align}
where we have used the relation in Eq.~(\ref{id1}).
Eq.~(\ref{ft0}) is exactly equal to the first term in Eq.~(\ref{f_intra_omega}) because 
$\omega_n$ and $\Delta_+$ are renormalized in the same manner by a factor $\eta_n$. 
The first term in Eq.~(\ref{ft3}) coinsides with the last term in Eq.~(\ref{f_intra_omega}).
But the interband impurity scatterings give rise to the term proportional to $I$.
The gap equation results in,
\begin{align}
&\bar{\Delta}_\lambda =  T\sum_{\omega_n}^{\omega_c} \frac{\pi N_0}{|\omega_n|A_0} \nonumber\\
&\times\left[ 
\left\{ g_\lambda (\omega_n^2+\gamma^2) + \frac{v^2}{2}(g_\lambda+|g_{12}|) - \frac{I}{2}(g_\lambda-|g_{12}|) \right\}
 \bar{\Delta}_\lambda \right.\nonumber\\
&\left. +
\left\{ |g_{12}|(\omega_n^2+\gamma^2) + \frac{v^2}{2}(g_\lambda+|g_{12}|) + \frac{I}{2}(g_\lambda-|g_{12}|) \right\}
 \bar{\Delta}_{\bar{\lambda}} 
\right].\label{scf1}
\end{align}
The first terms in Eqs.~(\ref{ft0}) and (\ref{ft3}) recover the gap equation in the clean 
limit. 
By comparing Eq.~(\ref{scf1}) with Eq.~(\ref{scf0}), the effects of impurity scatterings
are represented by $I$ which is derived from the interband impurity scatterings.
The pair density suppressed by the interband
impurity scatterings explains the physical meaning of $I$ which is originated from $s_a$ in Eq.~(\ref{sa_def}) 
through the second term in Eq.~(\ref{self1}). 
As shown in Eq.~(\ref{f_intra_omega}), $s_a$ is proportional to the pairing correlations after summing over 
$\boldsymbol{k}$,
\begin{align}
\hat{s}_a = & \frac{1}{\pi N_0} \left[ \langle f_2 \rangle \hat{\rho}_2 - \langle f_3 \rangle \hat{\rho}_3 \right].
\end{align}
At the last term of Eq.~(\ref{ft3}), $\langle f_2 \rangle$ couples $f_2$ and $\langle f_3 \rangle$ couples $f_3$.
The summation over $\boldsymbol{k}$ with the renormalized frequency $\tilde{\omega}$ gives $I$ as
\begin{align}
I\propto -\langle f_2 \rangle\langle f_2 \rangle_{\omega_n\to\tilde{\omega}_n} + 
\langle f_3 \rangle \langle f_3 \rangle_{\omega_n\to \tilde{\omega}_n}.
\end{align}
Therefore $I$ is proportional to the pair density that are removed by the interband impurity scatterings.
The suppression of the even- (odd-) frequency pairing correlation decreases (increases) $T_c$.
The odd-frequency symmetry of a Cooper pair accounts for the negative sign of the first term in $I$.
We note in the case of $\bar{\Delta}_1=\bar{\Delta}_2$ that $T_c$ remains unchanged from its value in the clean limit because
of $s_a=0$. This conclusion agrees with the results in the previous papers~\cite{golubov:prb1997,efremov:prb2011}.

\begin{figure}[tbh]
\begin{center}
\includegraphics[width=8.5cm]{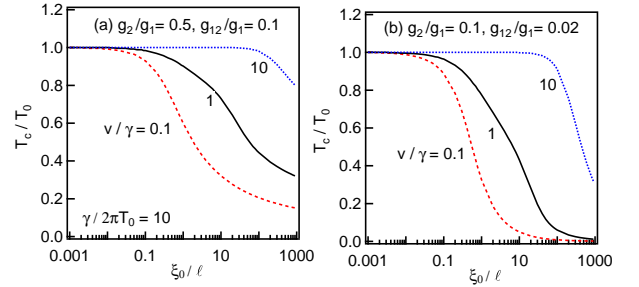}
\end{center}
\caption{The superconducting transition temperature in a two-band superconductor 
is plotted as a function of $\xi_0/\ell$ for $g_2/g_1=0.5$ in (a) and $g_2/g_1=0.1$ in (b).
The vertical axis is normalized to the transition temperature in the clean limit $T_0$.
We fix the band asymmetry at $\gamma/(2\pi T_0)=10$ and the ratio of $g_{12}/g_2=0.2$ in both (a) and (b).}
\label{fig:tc}
\end{figure}

In Fig.~\ref{fig:tc}, we plot the transition temperature $T_c$ as a function of $\xi_0/\ell$ 
for $g_2/g_1=0.5$ in (a) and $g_2/g_1=0.1$ in (b), where $T_0$ is the transition temperature 
in the clean limit, $\xi_0 =  v_F/(2\pi T_0)$, $\ell=v_F \tau_{\mathrm{imp}}$ and $g_{12}=0.2\, g_2$. 
All the results show that the transition temperature decreases with the 
decrease of $\xi_0/\ell >1$, which can be explained by $I>0$ in Eq.~(\ref{i_def}).
The first term in Eq.~(\ref{i_def}) is smaller than the 
second term in all the parameter region, which leads to the suppression of
$T_c$. The results are consistent with those in the previous papers~\cite{golubov:prb1997,efremov:prb2011}.
The degree of $T_c$ suppression is smaller for larger $v/\gamma$. 
In the clean limit, the amplitude of the odd-frequency pairing correlation is proportional to $v$.
The negative sign of the first term in Eq.~(\ref{i_def}) reflects the fact that 
impurities break such odd-frequency pairs and stabilize the superconducting state. 
In a dirty regime at $\xi_0/\ell=10$ for example, $T_c$ increases with the increase of $v/\gamma$.
In experiments, the pressurizing a superconductor may modify the parameter $v/\gamma$. 
Therefore the presence of odd-frequency pairs affects the variation of $T_c$ of a dirty 
two-band superconductor under the physical pressure.   

\begin{figure}[tbh]
\begin{center}
\includegraphics[width=8.5cm]{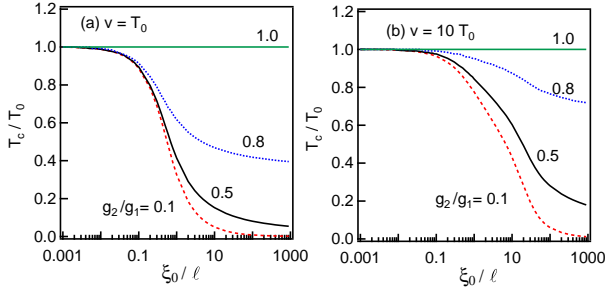}
\end{center}
\caption{The superconducting transition temperature in a two-band superconductor 
is plotted as a function of $\xi_0/\ell$ for several choices of $g_2/g_1$.
We choose $v=T_0$ in (a) and $v=10 T_0$ in (b).
We fix the ratio of $g_{12}/g_1 = 0.02$ and $\gamma =10\, T_0 $ in both (a) and (b).}
\label{fig:tc2}
\end{figure}

In Fig.~\ref{fig:tc2}, we plot the transition temperature $T_c$ as a function of $\xi_0/\ell$ 
for $v= T_0$ in (a) and $v= 10 T_0$ in (b). 
We fix $g_{12}/g_1$ at 0.02 and $\gamma$ at $10 T_0$.  
The transition temperature decreases with the increase of $\xi_0/\ell >1$ for $g_2<g_1$.
The suppression of $T_c$ become weaker as $g_2/g_1$ goes unity. 
The gap equation at $g_1=g_2$ always gives rise to a solution of $\bar{\Delta}_1 =\bar{\Delta}_2$. As a result, 
 $s_a$ in Eq.~(\ref{sa_def}) vanishes identically because of $\Delta_-=\bar{\Delta}_1-\bar{\Delta}_2=0$.
Therefore, $T_c$ is independent of $\xi_0/\ell$ in the symmetric case~\cite{golubov:prb1997,efremov:prb2011}. 

The suppression of $T_c$ in a dirty two-band $s$-wave superconductor
is not analogus to that of an unconventional superconductor in the presence of impurities. 
To make the difference clear, we consider
the gap equation in the dirty limit $\mu_F \gg 1/\tau_{\mathrm{imp}} \gg T_c, \gamma$ and $v$.
Here we assume $\theta=0$, $g_1>g_2 \gg g_{12}>0$, and $v=0$ for simplicity.
In the dirty limit, $I$ in Eq.~(50) goes to $\omega_n^2+\gamma^2$. 
The resulting gap equation in Eq.~(53) is given by
\begin{align}
\bar{\Delta}_1 =&  (g_1 +g_{12})\, N_0\, J_T  (\bar{\Delta}_1 + \bar{\Delta}_2),\\
\bar{\Delta}_2 =&  (g_2 +g_{12})\, N_0\,  J_T  (\bar{\Delta}_1 + \bar{\Delta}_2),\\
J_T=& 2\pi T \sum_{\omega_n>0}^{\omega_c}\frac{1}{\omega_n}.
\end{align}
The solution of the equation exists when 
\begin{align}
\frac{({g}_1+{g}_2)}{2}\, N_0\,  J_{T_c}=1, \label{gap_dirty}
\end{align}
 is satisfied.
In the clean limit, the gap equation is given by, 
\begin{align}
{g}_1 J_{T_0}\, N_0 =1, \label{gap_clean}
\end{align}
with $T_0$ being the transition temperature in the clean limit.
The attractive interaction in the clean limit $g_1$ 
decreases effectively to $(g_1+g_2)/2$ in the dirty limit. 
Therefore,
$T_c$ obtained from Eq.~(\ref{gap_dirty}) becomes smaller when the asymmetry between $g_1$ and $g_2$ 
is larger. This analysis explains well the numerical results in Fig.~\ref{fig:tc2}. 
On the other hand in unconventional superconductors characterized by such symmetry as $p$- and $d$-wave, 
the gap equation in the presence of impurity scatterings is given by
\begin{align}
1 = g\, N_0\, 2\, \pi\, T\sum_{\omega_n>0}^{\omega_c} 
\left. \frac{1}{\omega_n + 1/(2\tau_{\mathrm{imp}})}\right|_{T=T_c}.
\end{align}
The impurity scatterings remove the singularity at the denominator, which 
leads to the strong suppression of $T_c$. 
As a result, $T_c$ goes to zero around $\xi_0/\ell =0.28$.

\section{$\Delta_1 \Delta_2<0$ and TRS-breaking states}
%

In Sec.~\ref{impurity}, we have discussed the effects of impurity scatterings on $T_c$ for 
$\bar{\Delta}_1 \bar{\Delta}_2>0$ called as $s_{++}$ state in recent literature. 
Here we briefly discuss a case of $\bar{\Delta}_1 \bar{\Delta}_2<0$ called as $s_{+-}$ state.
In our microscopic model, an $s_{+-}$ state is realized by choosing the phase 
as $\theta =\pi/2$ in Eqs.~(\ref{h0}) and (\ref{vimp}) and is unitary equivalent to the $s_{++}$ 
state as mentioned in Sec.~II.  
Therefore, the dependence of $T_c$ on $\xi_0/\ell$ for an $s_{+-}$ state is exactly the same 
as that for an $s_{++}$ state shown in Fig.~\ref{fig:tc}. 

\begin{figure}[tbh]
\begin{center}
\includegraphics[width=8.5cm]{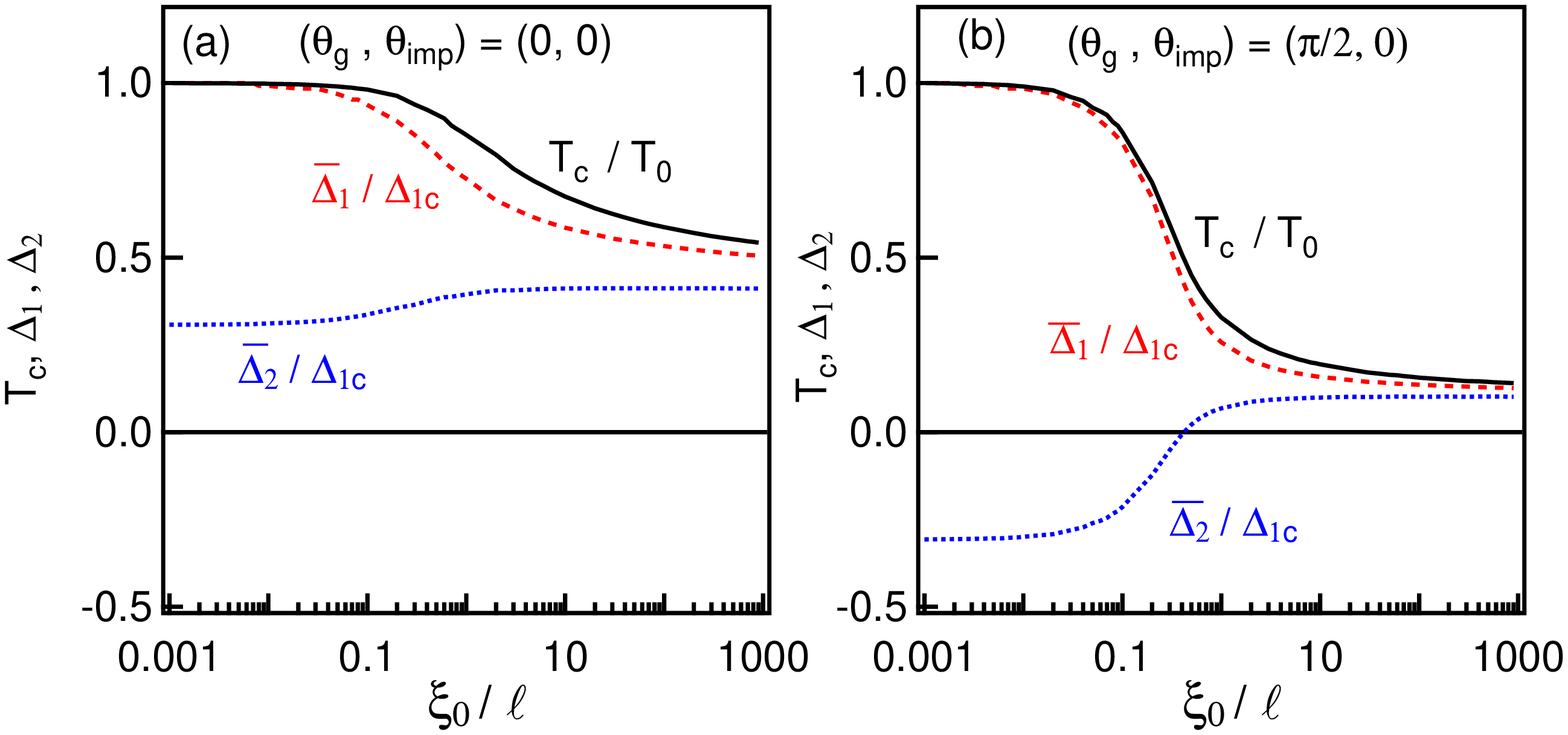}
\includegraphics[width=8.5cm]{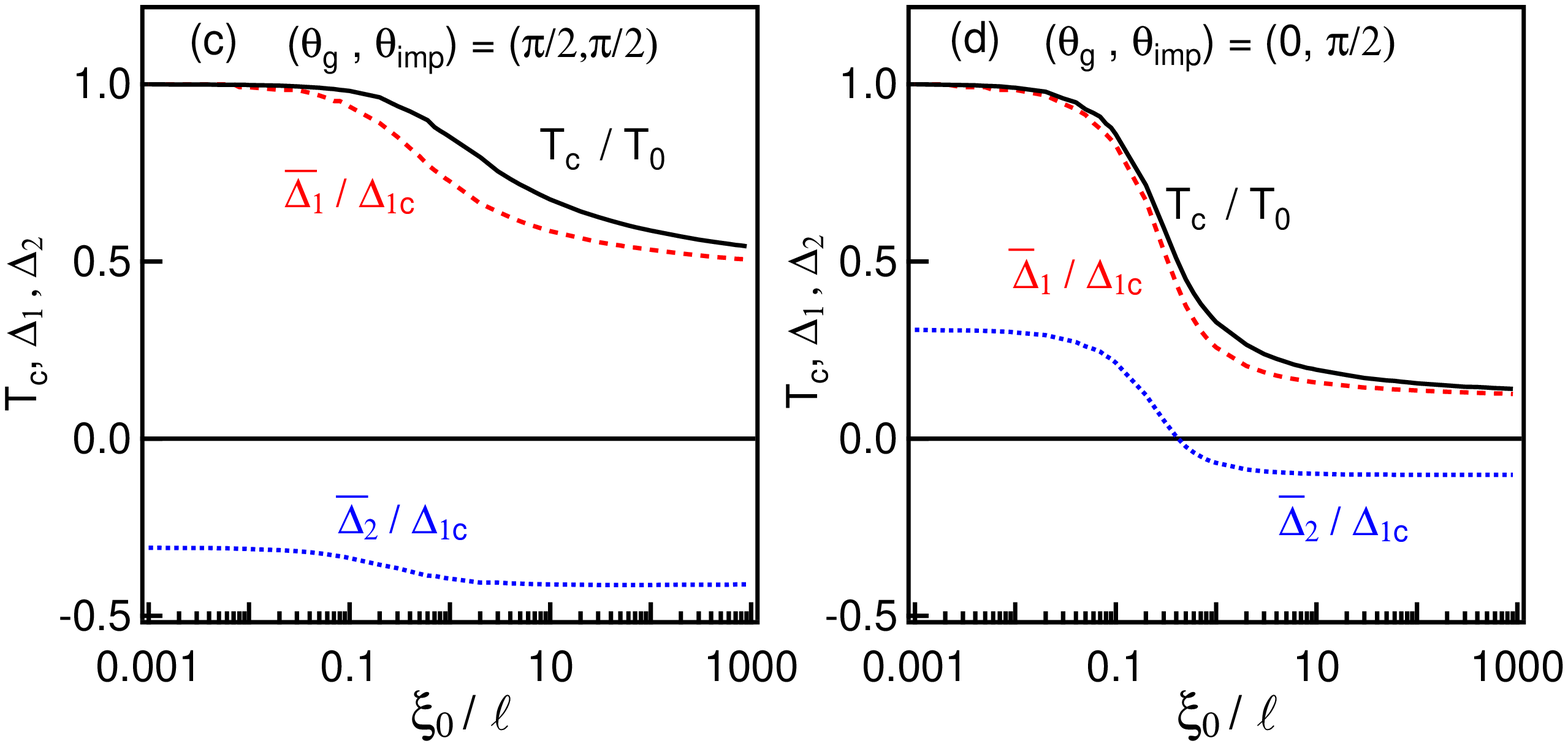}
\end{center}
\caption{The transition temperature ($T_c$) and the pair potentials ($\bar{\Delta}_1$, $\bar{\Delta}_2$) 
for $v=\gamma=0$, $g_2=0.8 g_1$ and $|g_{12}|=0.05 g_1$. 
The pair potentials are calculated at $T=0.5T_c$, where $\Delta_{1c}$ is the amplitude of $\bar{\Delta}_1$ in the clean 
limit. 
We introduce the phases of two potentials as $g_{12}=|g_{12}| e^{i \theta_g}$ and 
that at $V_{\mathrm{imp}}\, e^{i\theta_{\mathrm{imp}}}$. 
Time-reversal symmetry is preserved at $\theta_g=\theta_{\mathrm{imp}}$ in (a) and (c).
whereas it is broken for $\theta_g\neq \theta_{\mathrm{imp}}$ in (b) and (d).
}
\label{fig4}
\end{figure}

Finally, to make clear a relation between the present results and the results in the previous 
papers~\cite{efremov:prb2011,stanev:prb2014},
we discuss superconducting states described by the Hamiltonian in the absence of time-reversal symmetry.
In what follows, we delete the hybridization and the asymmetry in the two bands for simplicity, (i.e., $v=\gamma=0$).
We also introduce two phases
\begin{align}
g_{12}=|g_{12}|e^{2\, i\, \theta_g}, \quad V_{\mathrm{imp}}\, e^{i\, \theta_{\mathrm{imp}}},\label{2phase}
\end{align}
where $\theta_g$ is the phase of interaction in Eq.~(\ref{bd_def}) and $\theta_{\mathrm{imp}}$ 
is the phase of the interband impurity potential in Eq.~(\ref{vimp}). 
They must be equal to each other and satisfy Eq.~(\ref{trs_cond}) to preserve time-reversal symmetry.
Here we choose $\theta_g$ and $\theta_{\mathrm{imp}}$ 
are either $0$ or $\pi/2$ independently to demonstrate the transition between an $s_{++}$ state and an $s_{+-}$ state.
At first we consider the case of $\theta_{\mathrm{imp}}=0$. 
The gap equation are
represented by
\begin{align}
\bar{\Delta}_1= &\pi\, N_0\, T\sum_{\omega_n}\left[
\frac{g_1\, D_1 }{\sqrt{\omega_n^2+D_1^2}} 
+\frac{g_{12}\, D_2}{\sqrt{\omega_n^2+D_2^2}} \right], \label{gap61}\\
\bar{\Delta}_2= &\pi\, N_0\, T\sum_{\omega_n} \left[
\frac{g_2\, D_2}{\sqrt{\omega_n^2+D_2^2}} 
+\frac{g_{12}\, D_1}{\sqrt{\omega_n^2+D_1^2}} \right],\label{gap62}\\
\Omega_\lambda =& \sqrt{\omega_n^2+\Delta_\lambda^2}, \quad
\eta=1+\frac{1}{4\tau_{\mathrm{imp}}}
\left[ \frac{1}{\Omega_1} +\frac{1}{\Omega_2}\right]. 
\end{align}
Here the pair potentials are modified as
\begin{align}
D_1=&\bar{\Delta}_1-\frac{\bar{\Delta}_1-\bar{\Delta}_2}{4 \tau_{\mathrm{imp}}\, \eta\, \Omega_2},
\quad D_2=\bar{\Delta}_2+\frac{\bar{\Delta}_1-\bar{\Delta}_2}{4 \tau_{\mathrm{imp}}\, \eta\, \Omega_1}.
\end{align}
at $\theta_{\mathrm{imp}} =0$.
The numerical results of $T_c$, $\bar{\Delta}_1$ and $\bar{\Delta}_2$ are plotted as a function of 
$\xi_0/\ell$ in Fig.~\ref{fig4}(a) and (b), where we choose $g_2= 0.8\, g_1$ and $|g_{12}|=0.05\, g_1$.
The pair potentials are calculated at $T=0.5\, T_c$, where 
$\Delta_{1c}$ is the amplitude of $\bar{\Delta}_1$ in the clean limit. 
In Fig.~\ref{fig4}(a), we set $\theta_g=0$ so that time-reversal 
symmetry is preserved in the Hamiltonian. 
An $s_{++}$ state is realized in both the clean limit and the dirty limit. 
In Fig.~\ref{fig4}(b), however, we set $\theta_g=\pi/2$ to realize an $s_{+-}$ state in the clean limit. 
The pair potential $\bar{\Delta}_2$ changes its sign around $\xi_o/\ell=0.3$. 
The superconducting state undergoes the transition from an $s_{+-}$ state to an $s_{++}$ state 
due to the impurity scatterings. We note in this case that time-reversal 
symmetry is broken because of $\theta_{\mathrm{imp}} \neq \theta_g$.
The transition can be understood by the gap equations in linear regime
\begin{align}
\bar{\Delta}_\lambda = & T\sum_{\omega_n}^{\omega_c} \frac{\pi N_0}{|\omega_n|}
\left[ \right. \nonumber\\
+& \left\{ g_\lambda \left(1- \frac{1}{4\tau_{\mathrm{imp}}|\tilde{\omega}_n|}\right) + 
\frac{s_{v}\, g_{12}}{4\tau_{\mathrm{imp}}|\tilde{\omega}_n|} \right\}
 \bar{\Delta}_\lambda  \nonumber\\
 +&\left.
\left\{ g_{12} \left(1- \frac{1}{4\tau_{\mathrm{imp}}|\tilde{\omega}_n|}\right) 
 + \frac{s_{v}\, g_{\bar{\lambda}}}{4\tau_{\mathrm{imp}}|\tilde{\omega}_n|} \right\}
 \bar{\Delta}_{\bar{\lambda}} 
\right],\label{scf_ef}\\
s_v=& \left\{ \begin{array}{rl} 1 & :\theta_{\mathrm{imp}}=0 \\
-1 & :\theta_{\mathrm{imp}}=\pi/2
\end{array}\right..
\end{align}
%
The coefficient of $\bar{\Delta}_\lambda$ in the first line is always positive.
Since $g_{12}<0$ at $\theta_g=\pi/2$ in Eq.~(\ref{2phase}), 
the coefficient of $\bar{\Delta}_{\bar{\lambda}}$ in the second line is negative in the lean limit.
Namely an $s_{+-}$ state is stable in the clean limit. 
On the other hand in the dirty limit, the sign of the second line in
 Eq.~(\ref{scf_ef}) becomes positive because of $g_1 > g_2 \gg |g_{12}|$ 
 and $s_{v}=1$. 
As a result, the impurity scatterings stabilize an $s_{++}$ state as shown in Fig.~\ref{fig4}(b). 
These results, however, do not mean that an $s_{++}$ state is more robust than an $s_{+-}$ state.

Secondly we consider the case of $\theta_{\mathrm{imp}}=\pi/2$. 
The gap equations are given by Eqs.~(\ref{gap61}) and (\ref{gap62}) with
\begin{align}
D_1=&\bar{\Delta}_1-\frac{\bar{\Delta}_1+\bar{\Delta}_2}{4 \tau_{\mathrm{imp}}\, \eta\, \Omega_2},
\quad D_2=\bar{\Delta}_2-\frac{\bar{\Delta}_1+\bar{\Delta}_2}{4 \tau_{\mathrm{imp}}\, \eta\, \Omega_1}.
\end{align}
at $\theta_{\mathrm{imp}} =\pi/2$. The numerical results are shown in Fig.~\ref{fig4}(c) and (d).
In (c), we set $\theta_g=\pi/2$ to preserve time-reversal symmetry. An $s_{+-}$ state is always 
realized for all $\xi_0/\ell$. On the other hand,  
the numerical results for $\theta_g=0$ in (d) show the transition 
from an $s_{++}$ state to an $s_{+-}$ state by the impurity scatterings.
The transition can be described well by the gap equation in linear regime in Eq.~(\ref{scf_ef}) 
 with $s_{v}=-1$ at $\theta_{\mathrm{imp}}=\pi/2$.
At $\theta_g=0$, the coefficient in the second line in Eq.~(\ref{scf_ef}) is 
positive in the clean limit and changes its sign to negative in the dirty limit.

Time-reversal symmetry is preserved in $\check{H}_0$ as long as $\theta_g$ satisfies 
Eq.~(\ref{trs_cond}). It is clear that time-reversal symmetry is always preserved 
in $\check{H}_{\mathrm{imp}}$ for all $\theta_{\mathrm{imp}}$. 
In the combined 
Hamiltonian $\check{H}=\check{H}_0+\check{H}_{\mathrm{imp}}$, however, the time-reversal symmetry is broken 
for $\theta_{\mathrm{imp}} \neq \theta_g$. 
The impurity scatterings causes the transition between an $s_{++}$ state and an $s_{+-}$ state 
in the absence of time-reversal symmetry as shown in Fig.~\ref{fig4} (b) and (d).
A spontaneously time-reversal symmetry broken state at a low temperature far below $T_c$~\cite{stanev:prb2014} 
can also be derived from such phase choice. 

In Ref.~\onlinecite{efremov:prb2011}, the gap equations are derived on the basis of the 
Eliahberg formula, where 
the self-energy due to the impurity scatterings is described in a phenomenological way. 
As a result, it is not easy to discuss time-reversal symmetry of the superconducting state within their formula. 
In this paper, on the other hand, we show that the phase transition between an $s_{+-}$ state and an $s_{++}$ state 
can be reproduced by the Green function theory for the mean-field Hamiltonian. 
In such cases, however, we conclude that time-reversal symmetry is broken in the Hamiltonian.

\section{Conclusion}

We studied the effects of random nonmagnetic impurities on the transition temperature
$T_c$ of a two-band superconductor on the basis of the standard Green function theory of superconductivity.
We assume an equal-time spin-singlet $s$-wave pair potential in each conduction band 
and consider the band hybridization as well as the band asymmetry.
The effects of impurity scatterings are taken into account through the self-energy which is estimated 
within the Born approximation. The transition temperature is calculated by solving 
the linearized gap equations for the pair potentials.
We assume that a two-band superconductor preserves time-reversal symmetry in both the absence 
and the presence of impurities.
Since an $s_{+-}$ state and an $s_{++}$ state are unitary equivalent to each other,
the interband impurity scatterings decrease $T_c$ in the two states exactly in the same manner. 
The variation of $T_c$ as a function of the band hybridization is explained well by the pair 
density removed due to impurity scatterings.

\begin{acknowledgments}
The authors are grateful to A. V. Balatsky, Y. Tanaka, and Y. V. Fominov for useful discussions.
This work was supported by Topological Materials Science (Nos.~JP15H05852 and JP15K21717) and 
KAKENHI (No.~JP15H03525) from the Ministry of Education, Culture, Sports, Science and Technology (MEXT) of 
Japan, JSPS Core-to-Core Program (A. Advanced Research Networks), Japanese-Russian JSPS-RFBR project 
(Nos.~2717G8334b and 17-52-50080),
and by the Ministry of Education and Science of the Russian Federation
(Grant No.~14Y.26.31.0007).
\end{acknowledgments}

\appendix


\section{Hamiltonian of a two-band superconductor} \label{model}
Let us begin the description of a two-band superconductor 
with the Hamiltonian of an electron at an isolated 
hydrogen-like atom,
\begin{align}
h_a =& - \frac{ \nabla^2}{2m} + v_a(\boldsymbol{r}), \\
h_a \, \phi_\lambda(\boldsymbol{r}) =& \epsilon_\lambda \, \phi_\lambda(\boldsymbol{r}). 
\end{align}
A number of atoms configure a regular lattice in a solid.
Thus, the Hamiltonian of such an atomic lattice becomes
\begin{align}
h_{\mathrm{N}} =& - \frac{\nabla^2}{2m} + v_L(\boldsymbol{r}), \\
v_L(\boldsymbol{r})=& \sum_{n=1}^N v_a(\boldsymbol{r}-\boldsymbol{R}_n),
\end{align}
where $n$ labels an atom and $\boldsymbol{R}_n$ points an atomic site.
The Bloch wave can be described as
\begin{align}
\Phi_{\lambda,\boldsymbol{k}}(\boldsymbol{r}) = \frac{1}{\sqrt{N}}\sum_{n} e^{i\boldsymbol{k}\cdot\boldsymbol{R}_n} 
\phi_\lambda(\boldsymbol{r}-\boldsymbol{R}_n).
\end{align}
We assume the orthonormal property
\begin{align}
\int d\boldsymbol{r}\, \phi_\lambda^\ast(\boldsymbol{r}-\boldsymbol{R}_n)\, \phi_{\lambda'}(\boldsymbol{r}-\boldsymbol{R}_{n'})
=\delta_{\lambda,\lambda'}\, \delta_{n,n'}.
\end{align}
This enable us to show the orthonormality and completness of the Bloch wave,
%
%

In what follows, we extract the two orbital degree of freedom, (i.e., $\lambda=1, 2$) and 
shrink the Hilbert space.
The electron operator in such Hilbert space is defined as 
\begin{align}
\Psi(\boldsymbol{r}) =&  \sum_{\boldsymbol{k}} \sum_{\lambda=1,2} 
\psi_{\lambda,\boldsymbol{k}}\, \Phi_{\lambda,\boldsymbol{k}}(\boldsymbol{r}),\\
\Psi^\dagger(\boldsymbol{r}) =&  \sum_{\boldsymbol{k}} 
\sum_{\lambda=1,2} \psi^\dagger_{\lambda,\boldsymbol{k}}\, \Phi^\ast_{\lambda,\boldsymbol{k}}(\boldsymbol{r}). \label{Psi_def_app}
\end{align}
%
The single-particle Hamiltonian is then given by
\begin{align}
&\mathcal{H}_{\mathrm{N}} =\int d\boldsymbol{r} 
\, \Psi^\dagger(\boldsymbol{r}) \, h_N \, \Psi(\boldsymbol{r}),\\
&=  \sum_{\lambda,\lambda'}\sum_{\boldsymbol{k}, \boldsymbol{k}'} \frac{1}{N} \sum_{n,n'}
\psi^\dagger_{\lambda,\boldsymbol{k}}\, \psi_{\lambda',\boldsymbol{k}'}
e^{-i\boldsymbol{k}\cdot \boldsymbol{R}_n}\, e^{i\boldsymbol{k}'\cdot \boldsymbol{R}_{n'}}\nonumber\\ 
&\times t_{\lambda,\lambda'}(\boldsymbol{R}_n-\boldsymbol{R}_{n'})\\
&t_{\lambda,\lambda'}(\boldsymbol{R}_n-\boldsymbol{R}_{n'}) \equiv 
\int d\boldsymbol{r}\, 
\phi^\ast_{\lambda}(\boldsymbol{r}-\boldsymbol{R}_n)\, h_{\mathrm{N}} \,\phi_{\lambda'}(\boldsymbol{r}-\boldsymbol{R}_{n'}).
\end{align}
\begin{widetext}
At $\boldsymbol{R}_n=\boldsymbol{R}_{n'}$, we find
\begin{align}
t_{\lambda,\lambda'}(0) = &
\int d\boldsymbol{r}\, 
\phi^\ast_{\lambda}(\boldsymbol{r}-\boldsymbol{R}_n)\, \left[ -\frac{\nabla^2}{2m}+ v_a(\boldsymbol{r}-\boldsymbol{R}_{n} )
+\sum_{m\neq n}v_a(\boldsymbol{r}-\boldsymbol{R}_{m}) \right]
 \,\phi_{\lambda'}(\boldsymbol{r}-\boldsymbol{R}_{n})
= \epsilon_{\lambda}\, \delta_{\lambda,\lambda'} + E_{\lambda,\lambda'},\\
E_{\lambda,\lambda'}\equiv& 
\int d\boldsymbol{r}\, 
\phi^\ast_{\lambda}(\boldsymbol{r}-\boldsymbol{R}_n)\, 
\sum_{m\neq 0}v_a(\boldsymbol{r}-\boldsymbol{R}_{m})
 \,\phi_{\lambda'}(\boldsymbol{r}-\boldsymbol{R}_{n}). \label{ell_app}  
\end{align}
The diagonal term $\epsilon_\lambda+ E_{\lambda,\lambda}$ 
gives the on-site potential for the $\lambda$ th band and the off-diagonal term 
represents the hybridization due to the crystalline field.
For $\boldsymbol{R}_n\neq \boldsymbol{R}_{n'}$, $t_{\lambda,\lambda'}$ represents the hopping integral 
among neighboring atoms.
The Hamiltonian becomes
\begin{align}
\mathcal{H}_{\mathrm{N}} =& \frac{1}{N} \sum_{n,n'} \sum_{\lambda,\lambda'} \sum_{\boldsymbol{k}, \boldsymbol{k}'}
\psi^\dagger_{\lambda,\boldsymbol{k}}\, \psi_{\lambda',\boldsymbol{k}'}
\left[ e^{-i(\boldsymbol{k}-\boldsymbol{k}') \cdot \boldsymbol{R}_n} 
\left\{ \epsilon_{\lambda}\, \delta_{\lambda,\lambda'} + E_{\lambda,\lambda'} \right\} \delta_{n,n'}
+ e^{-i\boldsymbol{k} \cdot \boldsymbol{R}_n}\, e^{i\boldsymbol{k}' \cdot \boldsymbol{R}_{n'}}
t_{\lambda,\lambda'}(\boldsymbol{R}_n-\boldsymbol{R}_{n'}) \right]\nonumber\\
=& \sum_{\lambda,\lambda'}\sum_{\boldsymbol{k}} 
\psi^\dagger_{\lambda,\boldsymbol{k}}\, 
\left\{ \epsilon_{\lambda}\, \delta_{\lambda,\lambda'} + E_{\lambda,\lambda'} + 
\sum_{\boldsymbol{\rho}}t_{\lambda,\lambda'}(\boldsymbol{\rho})e^{i\boldsymbol{k}\cdot \boldsymbol{\rho}} \right\}
\psi_{\lambda',\boldsymbol{k}},\\
\frac{1}{N} \sum_{n,n'} &t_{\lambda,\lambda'} (\boldsymbol{R}_n-\boldsymbol{R}_{n'})\,
  e^{-i\boldsymbol{k} \cdot \boldsymbol{R}_n}\, e^{-i\boldsymbol{k}' \cdot \boldsymbol{R}_{n'}}
= \delta_{\boldsymbol{k},\boldsymbol{k}'} \sum_{\boldsymbol{\rho}} 
t_{\lambda,\lambda'}(\boldsymbol{\rho})e^{i\boldsymbol{k}\cdot \boldsymbol{\rho}},
\end{align}
with $\boldsymbol{\rho}= \boldsymbol{R}_n-\boldsymbol{R}_{n'}$.
The Hamiltonian is represented in the matrix form
\begin{align}
\mathcal{H}_{\mathrm{N}} = \sum_{\boldsymbol{k}} 
[ \psi_{1,\boldsymbol{k}}^\dagger,\, \psi_{2,\boldsymbol{k}}^\dagger]
\left[\begin{array}{cc}
\epsilon_1 + E_{1,1} + t_{11}(\boldsymbol{k}) & E_{1,2} + t_{12}(\boldsymbol{k}) \\
E^{\ast}_{1,2} + t^\ast_{12}(\boldsymbol{k}) &
\epsilon_2 + E_{2,2} + t_{22}(\boldsymbol{k})
\end{array}\right]
\left[ \begin{array}{c} \psi_{1,\boldsymbol{k}} \\ \psi_{2,\boldsymbol{k}} \end{array}\right]. \label{h0_app}
\end{align}
In the text, we represent
\begin{align}
\epsilon_\lambda + E_{\lambda,\lambda} + t_{\lambda,\lambda}(\boldsymbol{k})=\xi_\lambda(\boldsymbol{k}), \quad 
E_{1,2}  =  v e^{i\theta}, \label{v_def_app}
\end{align}
and neglect the interband hopping $t_{12}(\boldsymbol{k})$. The impurity potential 
hybridizing the two bands should have the same phase factor $e^{i\theta}$.
The phase of hybridization $\theta$ depends on the choice of the orbital function $\phi_\lambda$.
Therefore, such phase should not affect physical values in the normal state.


The attractive interaction between two electrons is described by the two-particle Hamiltonian,
\begin{align}
\mathcal{H}_I=-\frac{1}{2} \int d\boldsymbol{r} \int d\boldsymbol{r}' \sum_{\sigma,\sigma'}
\Psi_{\sigma'}^\dagger(\boldsymbol{r}')\, 
\Psi_{\sigma}^\dagger(\boldsymbol{r})\,
u(\boldsymbol{r}-\boldsymbol{r}') \,
\Psi_{\sigma}(\boldsymbol{r})\,
\Psi_{\sigma'}^\dagger(\boldsymbol{r}'),
\end{align}
where $\sigma=\uparrow$ or $\downarrow$ represents spin of an electron.
By substituting Eq.~(\ref{Psi_def_app}) into the Hamiltonian, we find
\begin{align}
\mathcal{H}_I =& -\sum_{\boldsymbol{k}_1-\boldsymbol{k}_4}\sum_{\lambda_1-\lambda_4}\sum_{\sigma,\sigma'}
\psi_{\lambda_1,\boldsymbol{k}_1,\sigma'}^\dagger\,
\psi_{\lambda_2,\boldsymbol{k}_2,\sigma}^\dagger\,
\psi_{\lambda_3,\boldsymbol{k}_3,\sigma}\,
\psi_{\lambda_4,\boldsymbol{k}_4,\sigma'}\,
\frac{1}{N^2}\sum_{n_1-n_4} e^{-i\boldsymbol{k}_1\cdot \boldsymbol{R}_{n_1}}\,
e^{-i\boldsymbol{k}_2\cdot \boldsymbol{R}_{n_2}}\,
e^{i\boldsymbol{k}_3\cdot \boldsymbol{R}_{n_3}}
e^{i\boldsymbol{k}_4\cdot \boldsymbol{R}_{n_4}} I_{\mathrm{int}},\\
I_{\mathrm{int}}=&
\frac{1}{2} \int d\boldsymbol{r} \int d\boldsymbol{r}'
\frac{1}{N}\sum_{\boldsymbol{q}} u_{\boldsymbol{q}} e^{i\boldsymbol{q}\cdot(\boldsymbol{r}-\boldsymbol{r}')}
\phi^\ast_{\lambda_1}(\boldsymbol{r}'-\boldsymbol{R}_1)\,
\phi^\ast_{\lambda_2}(\boldsymbol{r}-\boldsymbol{R}_2)\,
\phi_{\lambda_3}(\boldsymbol{r}-\boldsymbol{R}_3)\,
\phi_{\lambda_4}(\boldsymbol{r}'-\boldsymbol{R}_4).
\end{align}
The space integral is estimated as follows,
\begin{align}
\int d\boldsymbol{r} e^{i\boldsymbol{q}\cdot \boldsymbol{r}} \, \phi_{\lambda_2}^\ast(\boldsymbol{r}-\boldsymbol{R}_2) \,
\phi_{\lambda_3}^\ast(\boldsymbol{r}-\boldsymbol{R}_3)
=&e^{i\boldsymbol{q}\cdot \boldsymbol{R}_2} 
\int d\boldsymbol{r}' e^{i\boldsymbol{q}\cdot \boldsymbol{r}'} \,
\phi_{\lambda_2}^\ast(\boldsymbol{r}') \,
\phi_{\lambda_3}^\ast(\boldsymbol{r}'-\boldsymbol{R}_3+\boldsymbol{R}_2),\\
\approx &
e^{i\boldsymbol{q}\cdot \boldsymbol{R}_2} \, \delta_{\boldsymbol{R}_3,\boldsymbol{R}_2}\; 
B_{\lambda_2,\lambda_3}(\boldsymbol{q}), \\
B_{\lambda,\lambda'}(\boldsymbol{q}) \equiv &
\int d\boldsymbol{r} e^{i\boldsymbol{q}\cdot \boldsymbol{r}} \,
\phi_{\lambda}^\ast(\boldsymbol{r}) \,
\phi_{\lambda'}(\boldsymbol{r})=B_{\lambda',\lambda}^\ast(\boldsymbol{-q}). \label{bll_app}
\end{align}
%
Together with
\begin{align}
\int d\boldsymbol{r} e^{-i\boldsymbol{q}\cdot \boldsymbol{r}} \, \phi_{\lambda_1}^\ast(\boldsymbol{r}-\boldsymbol{R}_1) \,
\phi_{\lambda_4}(\boldsymbol{r}-\boldsymbol{R}_4)
=&e^{i\boldsymbol{q}\cdot \boldsymbol{R}_1} 
\delta_{\boldsymbol{R}_1,\boldsymbol{R}_4}\; B_{\lambda_1,\lambda_4}(\boldsymbol{-q}),
\end{align}
we find
\begin{align}
\mathcal{H}_I =& -\sum_{\boldsymbol{k}_3,\boldsymbol{k}_4}\sum_{\lambda_1-\lambda_4}
\sum_{\sigma,\sigma'}\frac{1}{N}\sum_{\boldsymbol{q}}
\psi_{\lambda_1,\boldsymbol{k}_4- \boldsymbol{q},\sigma'}^\dagger\,
\psi_{\lambda_2,\boldsymbol{k}_3+\boldsymbol{q},\sigma}^\dagger\,
\psi_{\lambda_3,\boldsymbol{k}_3,\sigma}\,
\psi_{\lambda_4,\boldsymbol{k}_4,\sigma'}\,
 u_{\boldsymbol{q}} \, B_{\lambda_1,\lambda_4}(\boldsymbol{-q}) \, B_{\lambda_2,\lambda_3}(\boldsymbol{q}).
\end{align}
To derive the pairing Hamiltonian, we assume
$\boldsymbol{k}=\boldsymbol{k}_3=-\boldsymbol{k}_4$, 
$\boldsymbol{k}'=\boldsymbol{k}_3+\boldsymbol{q} =-\boldsymbol{k}_4+\boldsymbol{q}$, 
and $\sigma'=\bar{\sigma}$. By considering the short range interaction, we delete $\boldsymbol{q}$ dependence of 
$u_{\boldsymbol{q}}$. 
The results become
\begin{align}
\mathcal{H}_I =& -\frac{1}{2}\frac{1}{N}\sum_{\boldsymbol{k},\boldsymbol{k}'}\sum_{\lambda_1-\lambda_4}\sum_{\sigma}
\psi_{\lambda_1,-\boldsymbol{k}',\bar{\sigma}}^\dagger\,
\psi_{\lambda_2,\boldsymbol{k}',\sigma}^\dagger\,
\psi_{\lambda_3,\boldsymbol{k},\sigma}\,
\psi_{\lambda_4,-\boldsymbol{k},\bar{\sigma}}\,
 u \, B^\ast_{\lambda_4,\lambda_1}(\boldsymbol{k}-\boldsymbol{k}') \, 
 B_{\lambda_2,\lambda_3}(\boldsymbol{k}-\boldsymbol{k}').
\end{align}
We consider only the intraband pairing order parameter, which leads to $\lambda_1=\lambda_2=\lambda$ and $\lambda_3=\lambda_4=\lambda'$.
The pairing interaction between two electrons in the $\lambda$-th band is described by
\begin{align}
g_\lambda= u \, B^\ast_{\lambda,\lambda}(\boldsymbol{k}-\boldsymbol{k}') \, 
 B_{\lambda,\lambda}(\boldsymbol{k}-\boldsymbol{k}')
\end{align}
for $\lambda=1,2$. By the definition, $g_\lambda$ is a real number.
The matrix elements
\begin{align}
g_{12}=& u \, B^\ast_{2,1}(\boldsymbol{k}-\boldsymbol{k}') \, 
 B_{1,2}(\boldsymbol{k}-\boldsymbol{k}'),\quad
g_{21}=  u \, B^\ast_{1,2}(\boldsymbol{k}-\boldsymbol{k}') \, 
 B_{2,1}(\boldsymbol{k}-\boldsymbol{k}')=g_{12}^\ast.
\end{align}
represent the scattering of a Cooper pair at the first band to that at the second band.
The last equation hold true because $\boldsymbol{k}$ and $\boldsymbol{k}'$ are running argument.
Hereafter, we remove $\boldsymbol{k}-\boldsymbol{k}'$ dependence from $g_1$, $g_2$ and $g_{12}$
for simplicity. The two order parameters are defined by, 
\begin{align}
\Delta_1= g_1 \frac{1}{N}\sum_{\boldsymbol{k}}\langle \psi_{1,\boldsymbol{k},\uparrow}\,
\psi_{1,-\boldsymbol{k},\downarrow}\rangle,\quad
\Delta_2= g_2 \frac{1}{N}\sum_{\boldsymbol{k}} \langle \psi_{2,\boldsymbol{k},\uparrow}\,
\psi_{2,-\boldsymbol{k},\downarrow}\rangle. \label{del_def_app}
\end{align}
By decoupling the interaction Hamiltonian, we obtain
the mean-field Hamiltonian,
\begin{align}
\mathcal{H}_{I}^{\mathrm{MF}}
= 
\left( \Delta_1^\ast + \frac{g_{12}^\ast}{g_2} \Delta_2^\ast \right) 
\sum_{\boldsymbol{k}} \psi_{1, \boldsymbol{k}, \uparrow}\, \psi_{1, -\boldsymbol{k}, \downarrow}
+
\left( \Delta_2^\ast + \frac{g_{12}}{g_1} \Delta_1^\ast \right) 
\sum_{\boldsymbol{k}} \psi_{2, \boldsymbol{k}, \uparrow}\, \psi_{2, -\boldsymbol{k}, \downarrow} \nonumber\\
+
\left( \Delta_1 + \frac{g_{12}}{g_2} \Delta_2\right) 
\sum_{\boldsymbol{k}}  \psi^\dagger_{1, -\boldsymbol{k}, \downarrow} \, \psi^\dagger_{1, \boldsymbol{k}, \uparrow}
+
\left( \Delta_2 + \frac{g_{12}^\ast}{g_1} \Delta_1 \right) 
\sum_{\boldsymbol{k}} \psi^\dagger_{2, -\boldsymbol{k}, \downarrow} \, \psi^\dagger_{2, \boldsymbol{k}, \uparrow}.\label{pair_app}
\end{align}
By combining the single-particle Hamiltonian in Eq.~(\ref{h0_app}) and the pairing Hamiltonian in Eq.~(\ref{pair_app}), 
the BCS Hamiltonian for a two-band superconductor is given by
\begin{align}
\mathcal{H}_{0} =& \mathcal{H}_{\mathrm{N}} +  \mathcal{H}_I^{\mathrm{MF}},\\
=&\sum_{\boldsymbol{k}}
[\psi^\dagger_{1, \boldsymbol{k}, \uparrow}, \psi^\dagger_{2, \boldsymbol{k}, \uparrow}, \psi_{1, -\boldsymbol{k}, \downarrow},
\psi_{2, -\boldsymbol{k}, \downarrow}] 
\left[ \begin{array}{cccc}
\xi_1(\boldsymbol{k}) & v e^{i\theta} & \bar{\Delta}_1 & 0 \\
v e^{-i\theta} & \xi_2(\boldsymbol{k}) & 0 & \bar{\Delta}_2 \\
\bar{\Delta}_1^\ast & 0 & -\xi_1(\boldsymbol{k}) & - v e^{-i\theta} \\
0 & \bar{\Delta}_2^\ast & -v e^{i\theta} & -\xi_2(\boldsymbol{k}) 
\end{array}\right]
\left[\begin{array}{c}
\psi_{1, \boldsymbol{k}, \uparrow}\\
\psi_{2, \boldsymbol{k}, \uparrow}\\
\psi^\dagger_{1, -\boldsymbol{k}, \downarrow}\\
\psi^\dagger_{2, -\boldsymbol{k}, \downarrow}
\end{array}\right],\\
\bar{\Delta}_1 =& \Delta_1 + \frac{g_{12}}{g_2} \Delta_2,\quad
\bar{\Delta}_2 = \Delta_2 + \frac{g_{12}^\ast}{g_1} \Delta_1, \label{td_def_app}
\end{align}
where we have assumed $\xi_\lambda(\boldsymbol{k})=\xi_l^\ast(-\boldsymbol{k})$. 
In the text, we represents the Hamiltonian in real space.
Although we defined the order parameters in Eq.~(\ref{del_def_app}), 
the renormalized pair potentials in Eq.~(\ref{td_def_app}) enter the Hamiltonian.
Therefore, $\bar{\Delta}_\lambda=|\bar{\Delta}_\lambda| e^{i{\varphi}_\lambda}$ 
determines the character of superconducting state.

The hybridization defined in Eq.~(\ref{v_def_app}) is a complex number. 
The phase of hybridization is derived from Eq.~(\ref{ell_app}). 
In this paper, we consider a simple case, where 
$\phi_1^\ast(\boldsymbol{r})\phi_2(\boldsymbol{r})$ is decomposed into 
$e^{i\theta} \times R_{12}(\boldsymbol{r})$ with $R_{12}$ being a real function.
Accordingly, Eq.~(\ref{bll_app}) is described by 
\begin{align}
B_{1,2}(\boldsymbol{q}) = &
\int d\boldsymbol{r} e^{i\boldsymbol{q}\cdot \boldsymbol{r}} \,
R_{12}(\boldsymbol{r}) e^{i\theta},\quad
B_{2,1}^\ast(\boldsymbol{q}) = 
\int d\boldsymbol{r} e^{-i\boldsymbol{q}\cdot \boldsymbol{r}} \,
R_{12}(\boldsymbol{r}) e^{i\theta}.
\end{align}
As a consequence, we obtain $g_{12}=|g_{12}|e^{2i\theta}$.
The phase of hybridization and the phase of $g_{12}$ are related to each other.
In addition, the phase of interband impurity potential must be $e^{i\theta}$. 
As we discuss in Sec.~\ref{trs} and \ref{gorkov}, the relation $2\theta - {\varphi}_1 + {\varphi}_2 =2 \pi n$ 
should be satisfied to preserve time-reversal symmetry. 
Otherwise, the Gor'kov equation and the gap equation do not have stable solutions.

\section{Self-energy}\label{selfenergy}
The Green function in the presence of impurity potential is calculated 
within the second order perturbation expansion with respect to the impurity potential,
\begin{align}
 \check{\mathcal{G}}(\boldsymbol{r}-\boldsymbol{r}') \approx &
\check{\mathcal{G}}_0(\boldsymbol{r}-\boldsymbol{r}')
+\int d\boldsymbol{r}_1 \check{\mathcal{G}}_0(\boldsymbol{r}-\boldsymbol{r}_1)\, \overline{\check{H}_{\mathrm{imp}}(\boldsymbol{r}_1)}\, 
\check{\mathcal{G}}(\boldsymbol{r}_1-\boldsymbol{r}') \nonumber\\
+&\int d\boldsymbol{r}_1 \int d\boldsymbol{r}_2 \, 
\check{\mathcal{G}}_0(\boldsymbol{r}-\boldsymbol{r}_1)\, \overline{\check{H}_{\mathrm{imp}}(\boldsymbol{r}_1)\, 
\check{\mathcal{G}}_0(\boldsymbol{r}_1-\boldsymbol{r}_2)\, \check{H}_{\mathrm{imp}}(\boldsymbol{r}_2)}\, 
\check{\mathcal{G}}(\boldsymbol{r}_2-\boldsymbol{r}').
\end{align}
By using the properties of impurity potential in Eqs.~(\ref{vimp_1}) and (\ref{vimp_2}), we obtain
\begin{align}
 \check{\mathcal{G}}(\boldsymbol{r}-\boldsymbol{r}') = & 
\check{\mathcal{G}}_0(\boldsymbol{r}-\boldsymbol{r}')
+n_{\mathrm{imp}}\, v_{\mathrm{imp}}^2 \int d\boldsymbol{r}_1 
\check{\mathcal{G}}_0(\boldsymbol{r}-\boldsymbol{r}_1)\,  \hat{\tau}_3 \,
\check{\mathcal{G}}_0(0) \, \hat{\tau}_3 \, 
\check{\mathcal{G}}(\boldsymbol{r}_1-\boldsymbol{r}') \nonumber\\
&+n_{\mathrm{imp}}\, v_{\mathrm{imp}}^2 \int d\boldsymbol{r}_1 
\check{\mathcal{G}}_0(\boldsymbol{r}-\boldsymbol{r}_1)  \, \hat{\rho}_1\, \hat{\tau}_3 \,
\check{\mathcal{G}}_0(0) \,  \hat{\rho}_1\, \hat{\tau}_3 \, 
\check{\mathcal{G}}(\boldsymbol{r}_1-\boldsymbol{r}').
\end{align}
The second and the third terms are derived from the intraband impurity potential and the interband 
impurity potential at $\theta=0$, respectively. 
The term proportional to $\hat{\rho}_1 \hat{\tau}_3 \check{\mathcal{G}}_0 \hat{\tau}_3$ 
and that proportional to $ \hat{\tau}_3 \check{\mathcal{G}}_0 \hat{\rho}_1 \hat{\tau}_3$
do not appear because the final state after applying the second order 
perturbation expansion should be identical to the initial state 
in the Born approximation. 
By applying the Fourier transformation, the Green function becomes
\begin{align}
 \check{\mathcal{G}}(\boldsymbol{k},\omega_n) = &
\check{\mathcal{G}}_0(\boldsymbol{k},\omega_n) + \check{\mathcal{G}}_0(\boldsymbol{k},\omega_n) 
\left[ \check{\Sigma}_a + \check{\Sigma}_b \right]\, \check{\mathcal{G}}(\boldsymbol{k},\omega_n),
\end{align}
where the self-energy within the Born approximation are defined in Eqs.~(\ref{sigma_a1}) and (\ref{sigma_b1}). 
Using the relation $\check{\mathrm{G}}_0(\boldsymbol{k}, \omega_n)^{-1}=i\omega_n - \check{H}_0(\boldsymbol{k})$, 
we reach Eq.~(\ref{gorkov1}).

For representing the impurities self-energy, the momentum summation of the Green functions is necessary, 
\begin{align}
&\frac{1}{V_{\mathrm{vol}}} \sum_{\boldsymbol{k}} \check{\mathcal{G}}_0(\boldsymbol{k},\omega_n)
=\left[
\begin{array}{cc} \hat{g}^{(0)}_{\omega_n} & \hat{f}^{(0)}_{\omega_n} \\ \{ \hat{f}^{(0)}_{\omega_n} \}^\ast & 
- \{ \hat{g}^{(0)}_{\omega_n} \}^\ast 
\end{array}
\right],\\
&\hat{g}^{(0)}_{\omega_n}=\frac{\pi N_0 }{ A_0 |\omega_n| }(-i\omega_n ) A_0 \hat{\rho}_0, \label{g_intra_omega}\\
&\hat{f}^{(0)}_{\omega_n} =\frac{\pi N_0}{A_0|\omega_n| }
\left[  -A_0 \Delta_+  \hat{\rho}_0  
+\gamma v \Delta_-  \hat{\rho}_1
+ \omega_n v \Delta_-  \hat{\rho}_2   
-(A_0-v^2) \Delta_-  \hat{\rho}_3 
\right] =\sum_{\nu=0}^3 \langle f_\nu \rangle \hat{\rho}_\nu. \label{f_intra_omega}
\end{align}
We calculate the summation of the Green function as
\begin{align}
\frac{1}{V_{\mathrm{vol}}} \sum_{\boldsymbol{k}} 
\frac{c_0+c_2\, \xi^2}{\xi^4+2\xi^2(\omega_n^2-\gamma^2-v^2) + A_0^2}
=
N_0 \int^{\infty}_{-\infty} d\xi \, 
\frac{c_0+c_2\, \xi^2}{\xi^4+2\xi^2(\omega_n^2-\gamma^2-v^2) + A_0^2}
=
\frac{\pi N_0}{2|\omega_n|A_0}(c_0+c_2 \, A_0)\label{k-int}
\end{align}
where $c_0$ and $c_2$ are numerical constant,
$N_0$ is the density of states at the Fermi level, and $A_0=\omega_n^2+\gamma^2+v^2$.

The self-energy due to the intraband impurity scattering does not change $T_c$. 
This conclusion can be confirmed by using the identity,
\begin{align}
& \frac{\gamma^2 v^2 - \omega_n \tilde{\omega}_n v^2 + (\omega_n^2+\gamma^2)(\tilde{\omega}_n^2+\gamma^2)}{2\tau_{\mathrm{imp}} |\omega_n| (\omega_n^2+ v^2+ \gamma^2)} 
+ (\tilde{\omega}_n^2+\gamma^2)
 = (A_0-v^2) \frac{\tilde{A} } {A_0 } \frac{\tilde{\omega}_n}{\omega_n} . \label{id1}
\end{align}


\section{Relation to pnictide superconductors}\label{pnictide}

We briefly explain the relation between the mean-field Hamiltonian in this paper and 
 superconductivity in pnictides.
The normal state Hamiltonian in pnictide 
is described in momentum space by
\begin{align}
\mathcal{H}_{\mathrm{N}} =& \sum_{\boldsymbol{k}, \sigma}
[ d_{x, \boldsymbol{k}, \sigma}^\dagger, d_{y, \boldsymbol{k}, \sigma}^\dagger]
\left[\begin{array}{cc}
\epsilon_{x}(\boldsymbol{k}) -\mu& \epsilon_{xy}(\boldsymbol{k}) \\
\epsilon_{xy}(\boldsymbol{k}) & \epsilon_{y}(\boldsymbol{k})-\mu
\end{array}\right]
\left[ \begin{array}{c} d_{x, \boldsymbol{k}, \sigma} \\ d_{y, \boldsymbol{k}, \sigma}
\end{array}\right], \label{c_hn}
\end{align}
where 
$\epsilon_{x}(\boldsymbol{k})$, $\epsilon_{y}(\boldsymbol{k})$, and $\epsilon_{xy}(\boldsymbol{k})$
represent the dispersion of two orbitals and the hybridization on the two-dimensional 
tight-binding model.
For example in Ref.~\onlinecite{raghu:prb2008}, they are given by
\begin{align}
\epsilon_{x}(\boldsymbol{k})=& -2 t_1 \cos k_x - 2t_2 \cos k_y -4t_3 \cos k_x\cos k_y, \\
\epsilon_{y}(\boldsymbol{k})=& -2 t_2 \cos k_x - 2t_1 \cos k_y -4t_3 \cos k_x\cos k_y, \\
\epsilon_{xy}(\boldsymbol{k})=& -4 t_4 \sin k_x \,\sin k_y,
\end{align}
where $t_1-t_4$ are the hopping amplitudes on the tight-binding lattice and are real numbers.
Before turning into superconducting state, 
we briefly mention the phase of hybridization.
Putting a phase $e^{i\pi/2}=i$ to the hybridization is described by a unitary 
transformation,
\begin{align}
\mathcal{H}_{\mathrm{N}} =& \sum_{\boldsymbol{k}, \sigma}
[ d_{x, \boldsymbol{k}, \sigma}^\dagger, d_{y, \boldsymbol{k}, \sigma}^\dagger]
\, \hat{u}_2\, \hat{u}_2^\dagger\, \left[\begin{array}{cc}
\epsilon_{x}(\boldsymbol{k}) -\mu& \epsilon_{xy}(\boldsymbol{k}) \\
\epsilon_{xy}(\boldsymbol{k}) & \epsilon_{y}(\boldsymbol{k})-\mu
\end{array}\right] \, \hat{u}_2\, \hat{u}_2^\dagger\,
\left[ \begin{array}{c} d_{x, \boldsymbol{k}, \sigma} \\ d_{y, \boldsymbol{k}, \sigma}
\end{array}\right],\\
=
& \sum_{\boldsymbol{k}, \sigma}
[ d_{x, \boldsymbol{k}, \sigma}^\dagger, i\, d_{y, \boldsymbol{k}, \sigma}^\dagger]
 \left[\begin{array}{cc}
\epsilon_{x}(\boldsymbol{k}) -\mu& i\, \epsilon_{xy}(\boldsymbol{k}) \\
-i\, \epsilon_{xy}(\boldsymbol{k}) & \epsilon_{y}(\boldsymbol{k})-\mu
\end{array}\right] 
\left[ \begin{array}{c} d_{x, \boldsymbol{k}, \sigma} \\ -i\, d_{y, \boldsymbol{k}, \sigma}
\end{array}\right],\\
\hat{u}_2=&\mathrm{diag}[1, i].
\end{align}
Therefore the phase $e^{i\pi/2}$ does not play any roles in the normal state.
It is clear that any physical values in the normal state do not depend on this phase.

To describe superconducting state in the weak coupling theory, we assume 
the spatially uniform spin-singlet $s$-wave pair potential in each band. 
Since $\epsilon_{x}(\boldsymbol{k})\neq \epsilon_{y}(-\boldsymbol{k})$, 
spatially uniform interband pair potential is absent~\cite{asano17interband}. 
Thus we define $\Delta_x$ for $\epsilon_x$ orbital and $\Delta_y$ for $\epsilon_y$.
The mean-field Hamiltonian becomes
\begin{align}
H_{\mathrm{S}}=& \sum_{\boldsymbol{k}} D_{\boldsymbol{k}}^\dagger 
\left[\begin{array}{cccc}
\epsilon_x(\boldsymbol{k})-\mu & \epsilon_{xy}(\boldsymbol{k}) & \Delta_x & 0 \\
\epsilon_{xy}(\boldsymbol{k}) & \epsilon_{y}(\boldsymbol{k})-\mu & 0 & \Delta_y  \\
 \Delta_x & 0 &  -\epsilon_{x}(\boldsymbol{-k}) +\mu & -\epsilon_{xy}(\boldsymbol{-k})  \\
 0 & \Delta_y  &  -\epsilon_{xy}(\boldsymbol{-k}) & -\epsilon_{y}(\boldsymbol{-k})+\mu
\end{array}\right] D_{\boldsymbol{k}}, \label{c_hs}\\
D_{\boldsymbol{k}}=& [ d_{x,\boldsymbol{k},\uparrow}, d_{y,\boldsymbol{k},\uparrow}, 
d_{x,\boldsymbol{-k},\downarrow}^\dagger, d_{y,\boldsymbol{-k},\downarrow}^\dagger]^\mathrm{T},
\end{align}
where T means the transpose of the matrix. We consider the Hamiltonian in a partial Nambu space
in which spin of an electron is $\uparrow$ and that of a hole is $\downarrow$.
As discussed in the text, the phase of the hybridization fixes 
the relative phase of $\Delta_x$ and $\Delta_y$ in the presence of time-reversal symmetry.
In Eq.~(\ref{c_hs}), we assume that $\Delta_x$ and $\Delta_y$ are real positive numbers.
The single particle Hamiltonian Eq.~(\ref{c_hn}) can be diagonalized as 
\begin{align}
\left[\begin{array}{cc}
\epsilon_{x}(\boldsymbol{k}) -\mu& \epsilon_{xy}(\boldsymbol{k}) \\
\epsilon_{xy}(\boldsymbol{k}) & \epsilon_{y}(\boldsymbol{k})-\mu
\end{array}\right] \hat{u}_{\boldsymbol{k}} = \hat{u}_{\boldsymbol{k}}
\left[\begin{array}{cc}
\epsilon_{1,\boldsymbol{k}} -\mu & 0 \\
0 & \epsilon_{2, \boldsymbol{k}}-\mu
\end{array}\right],
\end{align}
by a unitary matrix $\hat{u}_{\boldsymbol{k}}$ with
\begin{align}
\hat{u}_{\mathrm{k}}
=&\left[\begin{array}{cc} \alpha_{\boldsymbol{k}} & \beta_{\boldsymbol{k}} \\
\beta_{\boldsymbol{k}} & -\alpha_{\boldsymbol{k}}
 \end{array}
\right], 
\quad \alpha_{\boldsymbol{k}}= \sqrt{\frac{1}{2}\left(1+\frac{\epsilon_-}{\sqrt{\epsilon_-^2+\epsilon_{xy}^2}}\right)}, 
\quad \beta_{\boldsymbol{k}}= \sqrt{\frac{1}{2}\left(1-\frac{\epsilon_-}{\sqrt{\epsilon_-^2+\epsilon_{xy}^2}}\right)}
\frac{\epsilon_{xy}(\boldsymbol{k})}{|\epsilon_{xy}(\boldsymbol{k})|},\\
\epsilon_\pm =& \frac{ \epsilon_{x}(\boldsymbol{k}) \pm \epsilon_{y}(\boldsymbol{k}) }{2}, 
\quad \epsilon_{1,\boldsymbol{k}} = \epsilon_+ + \sqrt{\epsilon_-^2+\epsilon_{xy}^2},
\quad \epsilon_{2,\boldsymbol{k}} = \epsilon_+ - \sqrt{\epsilon_-^2+\epsilon_{xy}^2}.
\end{align}
 The Hamiltonian in superconducting state in Eq.~(\ref{c_hs}) can be transformed into 
\begin{align}
H_{\mathrm{S}}=& \sum_{\boldsymbol{k}} D_{\boldsymbol{k}}^\dagger\, \check{U}_{\boldsymbol{k}}
\, \check{U}_{\boldsymbol{k}}^\dagger
\left[\begin{array}{cccc}
\epsilon_x(\boldsymbol{k}) -\mu& \epsilon_{xy}(\boldsymbol{k}) & \Delta_x & 0 \\
\epsilon_{xy}(\boldsymbol{k}) & \epsilon_{y}(\boldsymbol{k})-\mu & 0 & \Delta_y  \\
 \Delta_x & 0 &  -\epsilon_{x}(\boldsymbol{-k}) +\mu& -\epsilon_{xy}(\boldsymbol{-k})  \\
 0 & \Delta_y  &  -\epsilon_{xy}(\boldsymbol{-k}) & -\epsilon_{y}(\boldsymbol{-k})+\mu
\end{array}\right] \check{U}_{\boldsymbol{k}}
\, \check{U}_{\boldsymbol{k}}^\dagger
D_{\boldsymbol{k}}, \\
=& \sum_{\boldsymbol{k}} 
C_{\boldsymbol{k}}^\dagger\,
\left[\begin{array}{cccc}
\epsilon_{1, \boldsymbol{k}} -\mu& 0 & \Delta_1 & 0 \\
0 & \epsilon_{2, \boldsymbol{k}} -\mu& 0 & \Delta_2  \\
 \Delta_1 & 0 &  -\epsilon_{1,-\boldsymbol{k}}+\mu & 0  \\
 0 & \Delta_2  &  0 & -\epsilon_{2, -\boldsymbol{k}}+\mu
\end{array}\right]
C_{\boldsymbol{k}},\label{c_spp}\\
C_{\boldsymbol{k}}=& \check{U}_{\boldsymbol{k}}^\dagger\, D_{\boldsymbol{k},\sigma},\\
\check{U}_{\boldsymbol{k}}=&
\left[\begin{array}{cc}
\hat{u}_{\boldsymbol{k}} & 0 \\
0 & \hat{u}^\ast_{-\boldsymbol{k}}\end{array}\right], \quad 
\left[\begin{array}{cc}
\Delta_1 & 0 \\
0 & \Delta_2\end{array}\right] = \hat{u}_{\boldsymbol{k}}^\dagger 
\left[\begin{array}{cc}
\Delta_x & 0 \\
0 & \Delta_y\end{array}\right] \hat{u}_{-\boldsymbol{k}}^\ast. 
\end{align}
We have assume that the pair potentials are band-diagonal even after the transformation. 
This assumption is justified when $\Delta_x=\Delta_y$ is satisfied.
Because of the symmetric band structures between $\epsilon_{x}$ and 
$\epsilon_{y}$,  $\Delta_x=\Delta_y$ can be a reasonable condition.
Under this condition, we also find $\Delta_1=\Delta_2$. 
As shown in the previous papers~\cite{golubov:prb1997,efremov:prb2011}, 
$T_c$ in such a symmetric superconductor is insensitive to the impurity concentration. 
This conclusion is derived not because the matrix Hamiltonian in Eq.~(\ref{c_spp}) with $\Delta_1=\Delta_2$
seems to descrive 
$s_{++}$ state but because the pair potentials are symmetric $\Delta_x=\Delta_y$.
To confirm the statement, 
we finally consider the unitary transformation described by a matrix
\begin{align}
\check{U}_4=\left[ \begin{array}{cc} \hat{u}_2 & 0 \\ 0 & \hat{u}_2^\ast \end{array}\right]
=\mathrm{diag}[1, i, 1, -i]. 
\end{align}
The Hamiltonian is transformed into 
\begin{align}
H_S=& \sum_{\boldsymbol{k}} 
C_{\boldsymbol{k}}^\dagger\, \check{U}_4\, \check{U}_4^\dagger
\left[\begin{array}{cccc}
\epsilon_{1, \boldsymbol{k}} -\mu& 0 & \Delta_1 & 0 \\
0 & \epsilon_{2, \boldsymbol{k}}-\mu & 0 & \Delta_2  \\
 \Delta_1 & 0 &  -\epsilon_{1,\boldsymbol{k}}+\mu & 0  \\
 0 & \Delta_2  &  0 & -\epsilon_{2, \boldsymbol{k}}+\mu
\end{array}\right]
\check{U}_4\, \check{U}_4^\dagger
C_{\boldsymbol{k}},\\
=& \sum_{\boldsymbol{k}} 
B_{\boldsymbol{k}}^\dagger\, 
\left[\begin{array}{cccc}
\epsilon_{1, \boldsymbol{k}}-\mu & 0 & \Delta_1 & 0 \\
0 & \epsilon_{2, \boldsymbol{k}} -\mu& 0 & -\Delta_2  \\
 \Delta_1 & 0 &  -\epsilon_{1,\boldsymbol{k}} +\mu& 0  \\
 0 & -\Delta_2  &  0 & -\epsilon_{2, \boldsymbol{k}}+\mu
\end{array}\right]
B_{\boldsymbol{k}},\label{c_spm}\\
B_{\boldsymbol{k}}=& \check{U}_{4}^\dagger\, C_{\boldsymbol{k}}.
\end{align}
The last matrix Hamiltonian with $\Delta_1=\Delta_2$ seems to describe $s_{+-}$ state.
The two matrix Hamiltonians in Eqs.~(\ref{c_spp}) and (\ref{c_spm}) are unitary 
equivalent to each other in the weak coupling theory. 
Therefore physical values derived from Eqs.~(\ref{c_spp}) 
are equal to those from Eqs.~(\ref{c_spm}). 
The argument above is valid even when we replace $i$ by $e^{i\theta}$ in $\hat{u}_2$.
When we assume two pair potentials $\Delta_x$ and $\Delta_y$ in the weak coupling theory, 
we immediately find that Eqs.~(\ref{c_spp}) and (\ref{c_spm})
are connected each other by the unitary transformation.

In a real material, a Cooper pair is formed by attractive electron-electron interactions 
which are mediated by bosons.
In the weak coupling theory, we usually integrate out such bosonic degree of freedom and define the 
pair potentials. This approximation enables us to have a mean-field Hamiltonian only for electrons. 
When the boson state is sensitive to the phase difference between the two order parameters, 
Eqs.~(\ref{c_spp}) and (\ref{c_spm}) are not unitary equivalent to each other.
To confirm this story, however, an expression of the interaction kernel or the effective electron-boson 
interaction Hamiltonian is necessary. The theory of superconductivity in this regime goes beyond 
the weak coupling limit.
A similar story is also possible when the phase difference between the two 
pair potentials couples to a gauge field. 
But these issues are beyond the scope of this paper.

 \end{widetext}


%

\end{document}